\documentclass[prb,aps,epsfig,longbibliography,twocolumn]{revtex4-1}
\usepackage{url,nicefrac,multirow,amssymb,amsfonts,tabularx, graphicx,amsmath,dcolumn,bm,color, colortbl, verbatim}
\usepackage[unicode=true,colorlinks=true]{hyperref}
\hypersetup{linkcolor=blue,citecolor=blue,urlcolor=blue}
\usepackage[dvipsnames]{xcolor}
\usepackage{mathrsfs}
\usepackage[titletoc,title]{appendix}

\newcommand*{\rom}[1]{\expandafter\@slowromancap\romannumeral #1@}

\newcommand{\ie}{\emph{i.e.}\,}
\newcommand{\ket}[1]{\left|#1\right\rangle }
\newcommand{\bra}[1]{\left\langle #1\right|}

\newcommand{\bsym}[1]{\boldsymbol{#1}}
\newcommand{\rbkt}[1]{\left( #1\right)}
\newcommand{\sbkt}[1]{\left[ #1\right]}

\newcommand{\spmtoi}[2]{t_{#1}\rbkt{\bsym{#2}}}
\newcommand{\tcm}[1]{t_{\text{c.m.}} \rbkt{\bsym{#1}}}
\newcommand{\trel}[2]{\tilde{t}_{#1}\rbkt{\bsym{#2}}}
\newcommand{\np}{N_{\phi}}
\newcommand{\jbkt}[2]{\left( #1 \middle| #2\right)}
\newcommand{\rtheta}[4]{\vartheta \sbkt{ \substack{{#1} \\ \\{#2}}}	\jbkt{#3}{#4}}

\newcommand{\iiota}{\dot{\iota}}
\def\mathclap#1{\text{\hbox to 0pt{\hss$\mathsurround=0pt#1$\hss}}}

\begin{document}
\title{Torus geometry eigenfunctions of an interacting multi-Landau level Hamiltonian}
\author{Abhishek Anand$^{1}$, Songyang Pu$^{2,3}$, and G. J. Sreejith$^{1}$}
\affiliation{$^1$Indian Institute of Science Education and Research, Pune 411008, India}
\affiliation{$^2$School of Physics and Astronomy, University of Leeds, Leeds LS2 9JT, United Kingdom}
\affiliation{$^3$Department of Physics, 104 Davey Lab, Pennsylvania State University, University Park, Pennsylvania 16802, USA}

\begin{abstract}
A short-ranged, rotationally symmetric multi-Landau-level model Hamiltonian for strongly interacting electrons in a magnetic field was proposed [A. Anand \textit{et al.}, Phys. Rev. Lett. \textbf{126}, 136601 (2021)] with the key feature that it allows exact many-body eigenfunctions on the disk not just for quasiholes but for all charged and neutral excitations of the entire Jain sequence filling fractions.
We extend this to geometries without full rotational symmetry, namely the torus and cylinder geometries, and present their spectra.
Exact diagonalization of the interaction on the torus produces the low-energy spectra at filling fraction $\nu=n/(2pn+1)$ that is identical, up to a topological $(2pn+1)$-fold multiplicity, to that of the integer quantum Hall spectra at $\nu=n$, for the incompressible state as well as all excitations.
While the ansatz eigenfunctions in the disk geometry cannot be generalized to closed geometries such as torus or sphere, we show how to extend them to cylinder geometry. Meanwhile, we show eigenfunctions for charged excitations at filling fractions between $\frac{1}{3}$ and $\frac{2}{5}$ can be written on the torus and the spherical geometries.
\end{abstract}
\maketitle

\section{Introduction} \label{intro}

Fractional Quantum Hall (FQH) systems exhibit a rich set of strongly interacting electronic phases.\cite{Klitzing80,Tsui82}
In spite of the strongly interacting nature of the FQH Hamiltonian, much progress has been made in understanding the phases due, in part, to the success of variational wave function approaches that describe the incompressible FQH ground states and their excitations. \cite{Laughlin83, Jain89, Jain89parton, Scarola2001, MOORE1991, RR99}
The ground states of the physical Coulomb interaction, which can be obtained through exact diagonalization (ED) in small systems contain complex electronic correlations. However the composite fermion (CF) variational wave functions accurately capture these correlations and are nearly identical to the Coulomb eigenfunctions. \cite{Jain89} The simple structure of the variational wave functions suggests the possibility of them being the exact ground states of simple Hamiltonians that may qualitatively resemble the physical interaction. This indeed turns out to be true for a subset of the variational states.

Haldane wrote a model short-range repulsive two-body interaction which has the Laughlin state as its unique densest zero-energy eigenfunction.\cite{Haldane83} Similarly a short range three-body repulsion produces the Moore-Read state as its ground state;\cite{MOORE1991} the three-body interaction has been argued to be generated by the perturbations induced by inter-Landau level scattering.\cite{Wojs10,Bishara09,Peterson13,Simon13,Sodemann13,Pakrouski15}
More general $n$-body interactions produce the Read-Rezayi sequence of states as the ground states.\cite{RR99,Simon07a}

A natural question is whether such parent Hamiltonians can be written for general composite fermion states. The CF wave functions describe the fractional quantum Hall effects (FQHE) at filling fractions $\nu=n/(2pn+1)$ where $n$ and $p$ are integers. 
The composite fermion wave functions for $N_e$ particles in flux $N_\phi$ have a general form $P_{\rm LLL}\mathcal{J}^{2p}(\{z_i\})\Phi(\{z_i\})$, where $\Phi(\{z_i\})$ is a Slater determinant of $N_e$ single particle Landau orbitals in flux $\np^*=N_\phi-2p(N_e-1)$ and $\mathcal{J}(\{z_i\})$ is the Jastrow factor. Here $z_i$ represents the coordinate of $i^{\rm th}$ particle. $P_{\rm LLL}$ projects the state into the lowest Landau level (LLL). The CF wave functions map the problem of interacting electrons in a magnetic field to that of more tractable non-interacting CFs in a reduced magnetic field. 
There has been recent success in constructing local two-body parent Hamiltonians for the entire set of unprojected CF states.\cite{Chen2017,Greiter2018,Bandyopadhyay18,Bandyopadhyay20,BoYang22}
Construction of such Hamiltonians for the projected CF states remains an open problem.\cite{Sreejith2018}

A strongly interacting model Hamiltonian can be constructed for states with a very similar structure namely $\mathcal{J}^{2p}(\{\hat{Z}_i\})\Phi(\{z_i\})$, where $\hat{Z}_i$ is the guiding center coordinate.\cite{Anand20} Although this does not solve the challenging problem of finding an exact Hamiltonian for the CF wave functions, the model has several appealing features.
As the electron density is changed, the same Hamiltonian produces incompressible states at Jain filling fractions of the form $n/(2pn+1)$ and allows exact eigenfunctions not just for the incompressible ground states and quasihole states but also the quasiparticle and neutral excitations. The low energy spectrum of the system at filling fraction $n/(2pn+1)$ has an exact one-to-one correspondence with the IQH states at integer filling factor $n$, just like what is seen in the actual problem. Berry phase and charge of localized excitations and entanglement spectrum of the ground state also match the ones expected from the standard CF theory for the Jain sequence fractions.
The model can be related to a multilayer system, with the different layers interpreted as different Landau levels of the same particle. Particles in every layer interact with other layers and within each layer without changing particle numbers in the layers. While the eigenstates in Fock space will be the same as that for a multilayer system, the real space eigenfunctions carry information of the distinct single particle states in different layers/LLs. Closely related ideas have been considered for Halperin wave functions.\cite{Papic2010, Goerbig2010} 
The underlying ideas can be successfully generalized to the case of non-Abelian Moore-Read states, where quasiholes, neutral excitations and quasiparticle states\cite{PhysRevLett.107.086806,Sreejith11,Sreejith13} become exactly solvable.\cite{Koji22}

In this work we focus on the Abelian cases, and complete the calculations in Ref.~\onlinecite{Anand20}, which focused on the disk and spherical geometries only, by studying the spectra and the eigenfunctions of the Hamiltonian in torus and cylinder geometries. We motivate the structure of the Hamiltonian in the torus using the analogy with the multilayer model. Direct generalization of disk eigenfunction to the torus is not possible as the function does not satisfy the necessary boundary conditions. Nonetheless, an alternative ansatz can be written for the torus and the sphere which describes the ground state and charged excitations at Laughlin filling fraction $1/(2p+1)$. For the case of cylinder, an ansatz can be constructed by direct generalization of the disk ansatz.

This paper is organized as follows. A brief overview of the model interaction on the disk geometry is provided in Sec.~\ref{SecModelInt}. In Sec.~\ref{SecSymTorus}, we review the many-body symmetries exclusive to the torus geometry. The model interaction for torus is given in Sec.~\ref{IntOnTorus}. 
In Sec.~\ref{ExactEig}, we extend the known disk eigenfunctions to the cylinder geometry. We show that, although these eigenfunctions do not extend to the torus, alternate ansatz eigenfunctions on torus can be constructed nevertheless for low-energy quasi-particle (QP) and some neutral excitations of $\nu=\frac{1}{3}$. 
Numerical results are presented in Sec.~\ref{NumRes} where various features of low-energy spectra of the model interaction for torus and cylinder are discussed. 
Finally, we summarize our results in Sec.~\ref{conclusion}.

\textit{Notations} -- In the paper, unless we mention otherwise, we use symmetric gauge $\bsym{A}(\bsym{r}) =\frac{B}{2}\bsym{r}\times \hat{e}_z$ corresponding to a magnetic field $\bsym{B}=-B\hat{e}_z$. The magnetic length is defined as $\ell=\sqrt{\hbar/eB}$ and the magnetic flux is counted in the units of flux quanta $\phi_0=hc/e$. When positions are represented in complex form, we use the convention $z=x+\iiota y$. For calculations in the disk geometry, angular momentum $k$ takes values in $0,1,2,\dots$ in all LLs. For given complex numbers $z$ and $\tau$, Jacobi theta function\cite{Mum83} is defined as
\begin{equation}
\rtheta{a}{b}{z}{\tau}= \sum_{n=-\infty}^{\infty}e^{\iiota \pi(n+a)^2\tau}e^{\iiota 2\pi(n+a)(z+b)} 
\end{equation}
where $a,\,b$ are rational numbers and $n$ is an integer.

\section{Model interaction in disk geometry} \label{SecModelInt}

In this section, we review the model interaction introduced in Ref.~\onlinecite{Anand20} and its ansatz eigenfunctions for the Jain sequence of filling fractions. We write the generic FQHE Hamiltonian in the disk geometry as
\begin{equation}
H = H_{\rm KE} + V \label{DiscHamiltonian}
\end{equation}
where $H_{\rm KE}$ represents the kinetic energy of the $N$ electrons with mass $m$ and charge $-e$ in a magnetic field, and is given by
\begin{equation}
H_{\rm KE} = \sum_{i}^N \frac{1}{2m}\rbkt{\bsym{p}_i + e\bsym{A}(\bsym{r}_i)}^2
\end{equation}
where $\bsym{p}_i$ is the canonical momentum of the $i$th electron which sees a perpendicular magnetic field given by $\text{B}\hat{e}_z= \nabla \times \bsym{A}(\bsym{r}_i)$.

The model interaction was motivated from the attempts at solving the open problem of constructing parent Hamiltonian for LLL projected CF wave functions. The CF wave function for $N$ electrons at filling $\nu=\nu^*/(2p\nu^*+1)$ on disk, is given by
\begin{equation}
\Psi^{\alpha}_{\nu=\frac{\nu^*}{2p\nu^*+1}} = \mathcal{P}_{\rm LLL} \prod_{i<j}^{N} ({z}_i - {z}_j)^{2p} \times	\Phi^{\alpha}_{\nu^*}  \label{CFstate}
\end{equation}
where $p$ is a positive integer, $\mathcal{P}_{\rm LLL} $ is the LLL projector and $z_i$ is position of the $i$th particle. $\Phi^{\alpha}_{\nu^*} $ is the Slater determinant state of single particle KE eigenstates.  The minimum relative angular momentum $m$ between a pair of CF in LL $n$ and $n'$ is given by $\delta_{n,n'}$.
Multiplication with the Jastrow factor $\mathcal{J}^{2p}(\{z\})=\prod_{i<j} ({z}_i - {z}_j)^{2p}$ increases the relative angular momentum of each pair of particles by $2p$.
This suggests that, for any pair of particles in LLs $n,\ n'$ in the Slater determinant $\Phi_{\nu^*} $, the relative angular momentum in the state $\Psi_{\nu}$ is forbidden in the range $\delta_{n,n'}, \delta_{n,n'} + 1,\dots,\delta_{n,n'}+2p$. One could consider a candidate exact Hamiltonian that imposes this constraint. However, the $\mathcal{J}^{2p}(\{z\})$ obfuscates the LL information and this approach fails. 

However, we can consider states of the following form:
\begin{equation}
\Psi^{\alpha}_{\nu=\frac{\nu^*}{2p\nu^*+1}} = \prod_{i<j}^{N} (\hat{Z}_i - \hat{Z}_j)^{2p} \times	\Phi^{\alpha}_{\nu^*}  \label{diskansatz}
\end{equation}
where the  position coordinates in the Jastrow factor are replaced with guiding-center coordinates $\hat{Z}_i$, which do not contain the LL scattering part. The modified Jastrow factor $\mathcal{J}^{2p}(\{\hat{Z}\})$ $(1)$ does not scatter particle into different LLs and $(2)$ still increases the relative angular momentum of each pair of particles by $2p$.

For this ansatz, the analysis presented before is valid and we can write a pseudopotential interaction $V$, given by
\begin{equation} \label{diskModelInter}
V= \sum_{n\leq n'=0}^{\infty}\;\sum_{m=\delta_{n,n'}}^{\delta_{n,n'}+2p-1} V^{n,n'}_m |n,n';m\rangle \langle n,n';m|.
\end{equation}
which penalizes only those relative angular momentum modes $m$ which are absent in the ansatz wave function. Here $|n,n';m\rangle$ is a two-particle state with relative angular momentum $m$ projected into Landau levels (LLs) $n$ and $n'$. The large positive numbers $V^{n,n'}_m$ represent energy penalties for specific relative momentum channels of the particle pairs. By construction, the ansatz wave functions defined in Eq.~\eqref{diskansatz} are zero-interaction energy (ZIE) eigenfunctions of this model interaction. Using extensive numerical tests, it was shown that states in the Eq.~\eqref{diskansatz} form a linearly independent complete basis for the ZIE eigenspace of the model interaction.

We work in a strong interaction limit which implies that all states outside the ZIE space are projected out to high energies. Since the kinetic energy $H_{\rm KE}$ commutes with $\hat{Z}$, the total energy [Eq.~\ref{DiscHamiltonian}] of the state is the same as the kinetic energy of the Slater determinant $\Phi^{\alpha}_{\nu^*} $. The ZIE degeneracy is lifted by $H_{\rm KE}$ and the low energy spectrum of this interacting system at filling fraction $\nu$ resembles the spectrum of the non-interacting IQH system at integer filling factor $\nu^*$.

Incompressible states at filling fraction $\nu$ correspond to the case where the Slater determinant $\Phi_{\nu^*}$ has an integer number $\nu^*$ of LLs compactly occupied. For finite systems, it was found that the ground state of the model and Coulomb interaction may be adiabatically connected. Its excited states (different excited states are labeled by $\alpha$) are constructed with the Slater determinants representing various excitations (quasiparticle/quasihole/neutral excitations) of this IQH state at $\nu^*$. Adding a particle to the incompressible configuration results in a quasiparticle excitation, whereas removing a particle from a fully filled LL generates a quasihole in the system. Neutral excitation is created when a particle in the fully filled LL jumps to a higher LL. A numerical study of adiabatic continuity of neutral excitations of model interaction to that of LLL projected Coulomb interaction is presented in Appendix \ref{NeutralModesAdiabatic}.  Figure \ref{fig:SlaterSchematic} shows representative LL occupations in the Slater determinant corresponding to each type of excitation.  

\begin{figure}[h]
	\includegraphics[width=0.9\columnwidth]{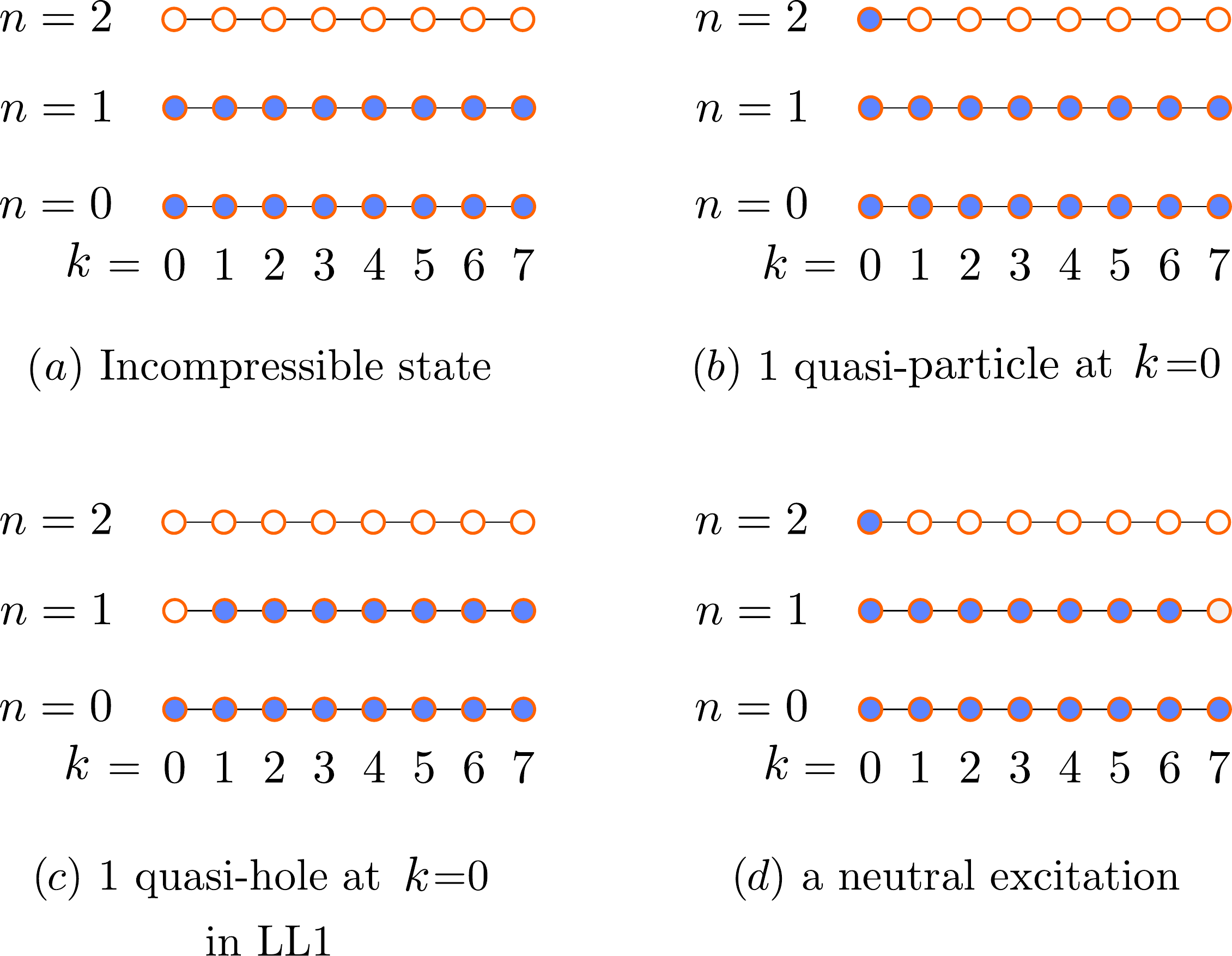}
	\caption{ Schematic representation of LL occupations of particles in the Slater determinant state $\Phi_{\nu^*}^{\alpha}$ (see Eq.~\ref{diskansatz}). Orange circles represent Landau orbitals whose LL indices and angular momentum are labeled by $n$ and $k$, respectively. Occupied orbitals are colored in blue.  $(a)$ shows an incompressible state at integer filling factor $\nu^*=2$, where particles completely fill the lowest two CF Landau-levels. LL occupation in $(b)-(d)$ are representative of $1$ quasi-particle, $1$ quasi-hole and $1$ neutral excitation at $\nu^*=2$, respectively. These excitations are labeled by $\alpha$ in Eq.~\eqref{diskansatz}.
	}
	\label{fig:SlaterSchematic}
\end{figure} 

Though the states (Eq.~\ref{diskansatz}) have a structure similar to the CF wave functions, they are qualitatively different. Unlike the CF wave functions projected to the LLL, these wave functions are distributed across LLs. The standard CF construction maps a Slater determinant wave function at an effective magnetic field $B^*$ to a state at magnetic field $B$ through multiplication by the Jastrow factor of particle coordinates. The Jastrow factor is defined such that the area of the droplet remains the same. Here, in Eq.~\ref{diskansatz}, the particles in the Slater determinant state $\Phi_{\nu^*}$ and the full state $\Psi$ experience the same magnetic field $B$, but the area of the droplet increases upon multiplication by the Jastrow factor of guiding-center coordinates. 

Nevertheless a key feature of the CF theory, which relates the spectrum of the interacting electrons to the spectrum of non-interacting composite fermions, is retained by the model interaction. This can also confirmed by exact diagonalization of the Hamiltonian in the case of the disk geometry. Although the ansatz eigenfunction is written only for the disk geometry, the exact diagonalization in the spherical geometry also shows the one-to-one correspondence with the IQH spectrum. 

In this work, we attempt to generalize the model and the exact eigenfunctions for disk geometry given in Eq.~\eqref{diskansatz} to the cylinder and torus geometries. 

\section{Symmetries on Torus} \label{SecSymTorus}
In this section we review the aspects of the torus geometry relevant to this work.
A torus represents a system with periodic-boundary conditions along lattice vectors $\bsym{L}_1$ and $\bsym{L}_2$. The periodicity implies that physical observables on torus must remain invariant under translations of type $\bsym{L}_{mn}= m\bsym{L}_1 + n\bsym{L}_2$, where $m,\,n$ are integers. 
In this section, we introduce the notations, and describe the conserved quantities of the many body states on the torus arising from the discrete symmetries. For a more detailed discussion, see Ref. \onlinecite{ManyBodyTranslationsTorus} and \onlinecite{BernevigTorus}.
\begin{figure}[h]
\includegraphics[width=0.6\columnwidth]{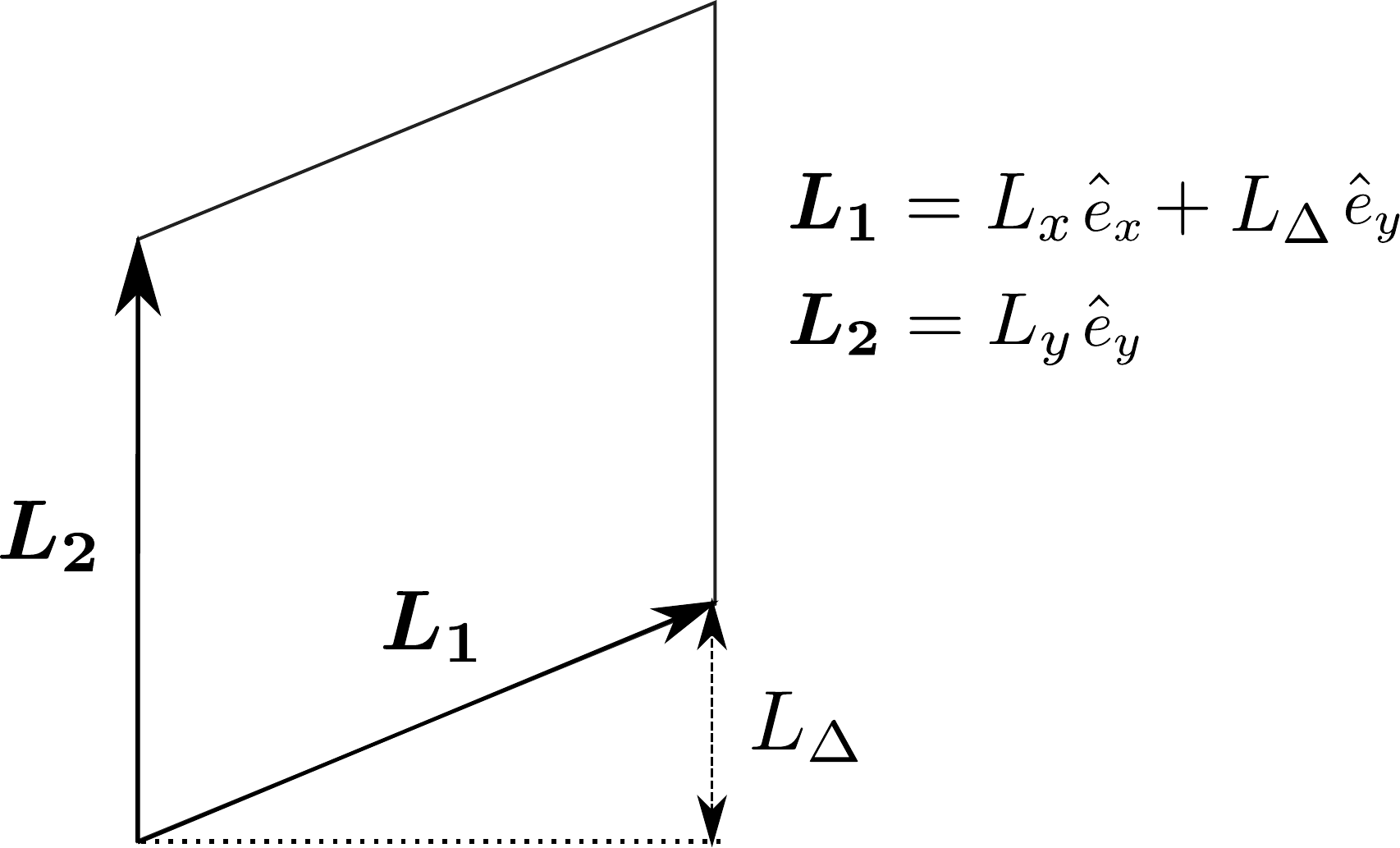}
\caption{
A general torus generated by $\bsym{L}_1$ and $\bsym{L}_2$. The skewness $L_{\Delta}$ parametrizes the deviation from a rectangular torus.
}
\label{fig::2.1}
\end{figure}

Hamiltonian for $N_e$ non-interacting electrons (with charge $-e$) in a uniform magnetic field $-{B}\hat{e}_z$ is given by 
\begin{equation}
H_{\text{\rm KE}} = \frac{1}{2m_e}\sum_{i}^{N_e} \bsym{\Pi}^{2}_i 
\label{Hfree}
\end{equation}
where, the kinetic momentum for the $i$th particle is given by $\bsym{\Pi}_i = \bsym{p}_i  + e \bsym{A}(\bsym{r}_i)$ and $\bsym{p} = - {\iiota}\hbar\bsym{\nabla}$ is the canonical momentum. The electron mass is $m_e$ and the gauge field $\bsym{A}(\bsym{r})$ satisfies $\nabla \times \bsym{A}(\bsym{r}) =-{B}\hat{e}_z$. In the presence of the magnetic field, usual spatial translations $T(\bsym{a})=e^{{\iiota}\bsym{a}\cdot \bsym{p}/\hbar}$ generated by canonical momentum do not commute with $H_{\rm free}$. A new operator, $\bsym{K}$, which commutes with $H_{\rm KE}$, is given by 
\begin{equation} \label{eq::2.1}
\bsym{K} = \bsym{\Pi} - \frac{\hbar}{\ell^2} \rbkt{\hat{e}_z\times \bsym{r}}
\end{equation}
 The usual translation operator, $T(\bsym{a})$, is replaced by magnetic translation operator (MTO), defined as  $t(\bsym{a})=e^{{\iiota}\bsym{a}\cdot \bsym{K}/\hbar}$.\cite{Brown64}  The lattice translation symmetries require that the state on the torus should remain invariant (up to a phase) under translations by the lattice vectors $t_i(\bsym{L}_1)$ and $t_i(\bsym{L}_2)$ for all $i$. It can be shown that this can happen only if the number of flux quanta through the unit cell --- given by $\np=\left|{\bsym{L}_1\times \bsym{L}_2}\right|/{2\pi\ell^2}$ --- is an integer.

All physical observables, including the many-body Hamiltonian $H = H_{\text{\rm KE}} + \sum_{i<j} V_{ij}$ will also be invariant under translations of type $t(\bsym{L}_{mn})$  on torus. Their Hilbert space representations are labeled by the eigenvalues $e^{\iiota \theta_1}$ and $e^{\iiota \theta_2}$ of $t_i(\bsym{L}_1)$ and $t_i(\bsym{L}_2)$, respectively. $\theta_1,\theta_2$ are identical for all particles.

Single particle eigenfunctions $\phi_{n,k}(z,\bar{z})$ of $H_{\text{free}}$, for the torus shown in Fig.~\ref{fig::2.1}, are given by
\begin{align}
\phi_{n,k}(z,\bar{z})&= \nonumber\\
&e^{-\frac{z^2+|z|^2}{2\ell^{2}}}\frac{(a_f^{\dagger})^{n}}{\sqrt{n}!}\sbkt{\rtheta{\frac{k}{\np}+\frac{\theta_2}{2\pi \np}}{\frac{\theta_{1}}{2\pi}}{\frac{\np z}{L_2}}{\np \tau}} \label{SingleParticleTorus}
\end{align}
where $0\leq k <\np$ is a $y$-momentum and $n=0,1,\dots$ is the LL-index. Here $z$ is the complex modular parameter, $\tau=-L_1/L_2 $  [where $L_1=L_x + \iiota L_{\Delta}$ and $L_2=\iiota L_y$] and $a_f^{\dagger} = \sqrt{2}\ell\rbkt{\frac{\bar{z}+z}{2\ell^2}-\partial_z} $ is the ladder operator for LL index such that its action on the exponential pre-factor is already taken into account. These states are eigenfunctions of translations $t_i(L_1)$ and $t_i(L_2)$, with eigenvalues $e^{\iiota \theta_1}$ and $e^{\iiota \theta_2}$, respectively. Many-body basis states can be written as $\ket{\{k_i\}}\equiv \ket{k_1,k_2,\dots,k_{N_e}}$ such that $ \ket{k} \equiv \phi_{n,k}(\bsym{r})$; the LL-index $n$ is suppressed for brevity. 

\begin{figure}
	\includegraphics[width=0.9\columnwidth]{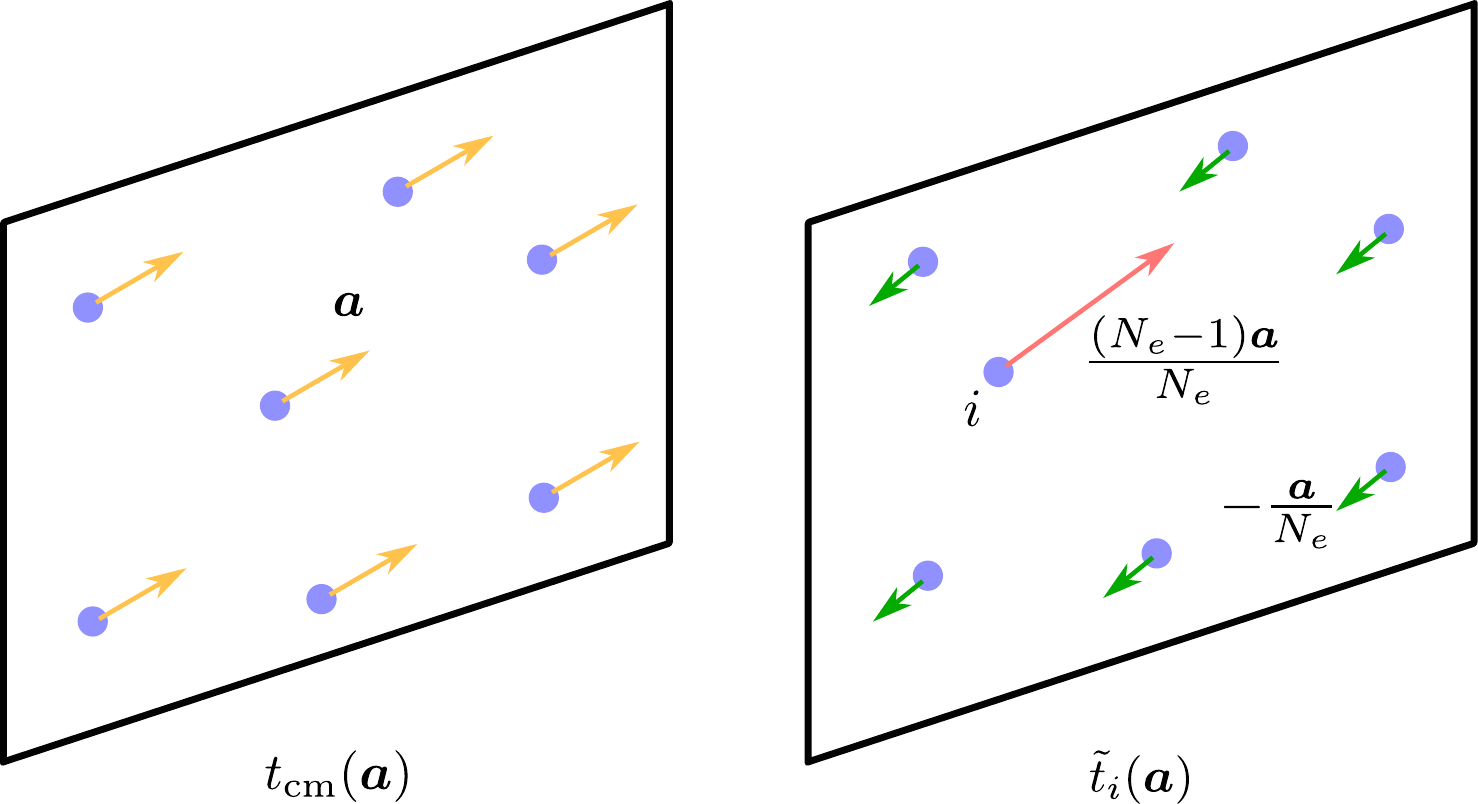}
	\caption{Schematic illustration for center-of-mass (left) and relative (right) many-body translations on torus, represented by $\tcm{a}$ and $\trel{i}{a}$, respectively. While $\tcm{a}$ translates all particles by same vector $\bsym{a}$, relative translation $\trel{i}{a}$ translates the $i$th particle by ${(N_e-1)\bsym{a}}/{N_e}$ and the rest are translated by $-{\bsym{a}}/{N_e}$.}
	\label{manyBodyTranslations}
\end{figure}

In Ref.~\onlinecite{ManyBodyTranslationsTorus}, it was  showed that, in addition to satisfying the boundary conditions of torus, eigenfunctions $\psi\rbkt{\{\bsym{r_i}\}}$ of a many-body Hamiltonian $H$ have additional conserved quantities which can be used to  label their spectra. These operators are given in the form of many-body translations, defined as

\begin{gather} \label{eq::2.4}
\trel{i}{a}={t}_{i}\rbkt{\frac{(N_e-1)\bsym{a}}{N_e}} \prod_{j\neq i} {t}_{j}\rbkt{-\frac{\bsym{a}}{N_e}}\nonumber \\
\tcm{a}= \prod_{i} \spmtoi{i}{a} 
\end{gather}
where ${\tilde{t}}_{i}$ and ${t}_{\rm c.m.}$ are named relative and center-of-mass translation operators respectively, schematically shown in Fig.~\ref{manyBodyTranslations}.

In order to preserve the Hilbert space representation defined by $(\theta_1,\theta_2)$, these operators need to commute with discrete lattice translations $t_j(\bsym{L}_1)$ and $t_j(\bsym{L}_2)$, for all $i,\, j$. Also, if we want the eigenfunctions to be simultaneously labeled with quantum numbers of  $\trel{i}{a}$ and $\tcm{a}$, these operators are also required to commute with each other, and among themselves for different translations. This can be achieved by forming a maximally commuting subset of these operators by restricting the translations. For a system of $N_e=pN$ particles in $\np = qN$ flux quanta, where $N=\text{gcd}(N_e,\np)$, this maximally commuting subset is given by
\begin{equation}\label{eq::2.3}
 \bigcup\limits_{m,n} \left\{
  \tilde{t}_{i}\rbkt{p\bsym{L}_{m,n}},  t_{\rm cm}\rbkt{\frac{q\bsym{L}_{m,n} + r\bsym{L}_{2}}{\np}}\;\middle|\;
   0 \leq r < q
\right\}
\end{equation}
Although this gives us infinite number of quantum numbers,  these quantum numbers  are fully determined by the eigenvalues of ${\tilde{t}}_{i}\rbkt{p\bsym{L}_1}$,  ${\tilde{t}}_{i}\rbkt{p\bsym{L}_2}$  and $t_{\rm cm}\rbkt{ \bsym{L}_{2}/\np}$. We calculate these eigenvalues below.

It is easy to check that many-body basis states $\ket{\{k_i\}}$ are already eigenfunctions of $t_{\rm cm}\rbkt{\bsym{L}_2/\np}$ with quantum number $K_2 = \sum_i k_i(\, \text{mod}\ \np)$, such that
\begin{equation}
t_{\rm cm}\rbkt{\bsym{L}_2/\np} \ket{\{k_j\},{K}_2} =  e^{{2\pi {\iiota}K_2/\np}}\ket{\{k_j\},{K}_2}
\end{equation}
If we denote a basis state in a given ${K}_2$-sector by $\ket{\{k_i\},{K}_2}$, the action of ${\tilde{t}}_{i}\rbkt{p\bsym{L}_1}$ is given by
\begin{gather}
{\tilde{t}}_i (p \bsym{L}_1) \ket{\{k_j\},{K}_2} = \ket{\{(k_j + q)\,\text{mod}\,\np \},{K}_2}\label{eq::2.5}
\end{gather}
Although ${\tilde{t}}_i (p \bsym{L}_1)$ increments $k_i$ for each particle by $q$, the state remains in the same $K_2$-sector. Eigenfunctions for ${\tilde{t}}_i (p \bsym{L}_1)$  are constructed by linearly combining states from a given $K_2$-sector as
\begin{gather} \label{eq::2.6}
\ket{\tilde{K}_1,{K}_2} = \sum_{j=0}^{N-1} e^{{\iiota}2\pi \tilde{K}_1 j/{N}} \ket{\{(k_i + jq)\,\text{mod}\,\np \},{K}_2}
\end{gather}
Eigenvalues of ${\tilde{t}}_i (p \bsym{L}_1)$ are $e^{-{\iiota}2\pi\tilde{K}_1/{N}}$ where $\tilde{K}_1$ can take values from $0,1,\dots,N-1$. 

We first exactly diagonalize the Hamiltonian in a fixed $K_2$-sector. In order to label the eigenfunctions with $\tilde{K}_1$ quantum numbers, we explicitly construct the ${\tilde{t}}_i (p \bsym{L}_1)$ operator using Eq.~\eqref{eq::2.5}, and compute its expectation value for the eigenstates. $\tilde{K}_1$ and $K_2$  can be  used to label the eigenfunctions of Hamiltonians which are invariant under lattice translations.

\section{Interaction on Torus}
\label{IntOnTorus}
The model interaction (Eq.~\ref{diskModelInter}) described in Ref.~[\onlinecite{Anand20}] was originally written in terms of pseudo-potentials for rotationally symmetric systems like a disk, which means that it assigns energy to a pair of particles based on their relative angular momentum given by $m$. The natural way to map the interaction into the torus geometry is to first reconstruct the real space form $V(|\bf{r}|)$ of the interaction  [or its Fourier transform $V(|\bf{q}|)$]. The interaction matrix elements on the torus can be calculated from $V(r)$. Unfortunately, the interaction shown in Eq.~\eqref{diskModelInter} is unlikely to be diagonal in the real-space representation. This is indicated by the fact that the projection of the interaction into each LL has the same number of non-zero psueodpotentials. 

We can nevertheless define torus matrix elements (between momentum states given in Eq.~\eqref{SingleParticleTorus}) of an interaction that produces the same features in the following way.  We construct a different real space form for different terms of the interaction in Eq.~\eqref{diskModelInter}. For instance, for $2p=2$, the interaction within the $n$th LL imposes an energy cost for the state $\left| n,n;m=1 \right\rangle = (a_1^\dagger)^n (a_2^\dagger)^n (b_1^\dagger-b_2^\dagger)\left|0,0\right\rangle$, which is a two particle state where both particles are in $n$th LL with relative momentum $m=1$, and $a_i$ and  $b_i$ are the $i$th-particle ladder operators for LL and  angular momentum respectively. We seek an interaction whose Fourier transform $V(|\bsym{q}|)$ satisfies
\begin{equation}
\int \text{d}\bsym{q}\, V(|\bsym{q}|) \langle n,n;m |e^{{\iiota}\bsym{q}\cdot {(\bsym{\hat{r}}_1-\bsym{\hat{r}}_2)}} | n,n;m \rangle   =\delta_{m,1}.
\end{equation}
 The expectation value $\langle n,n;m |e^{{\iiota}\bsym{q}\cdot {(\bsym{\hat{r}}_1-\bsym{\hat{r}}_2)}} | n,n;m \rangle$ can be evaluated to be $ \left\langle n |e^{A_1}|n \right\rangle  \left\langle n |e^{A_2}|n \right\rangle  \left\langle m| B|m \right\rangle $ where $\bsym{\hat{r}}_i$ are the position operators,  $A_i=\frac{\iiota \ell}{\sqrt{2}}(qa_i^\dagger + \bar{q}a_i)$,   $B=e^{(q\ell)^2/2}e^{\iiota\bar{q}\ell b_r^\dagger}e^{\iiota q\ell b_r}$ and $b_r=(b_1-b_2)/\sqrt{2}$ is the ladder operator for relative angular momentum. Here the reciprocal vector is written as a complex number $q=q_x+\iiota q_y$ (note that, this is unrelated from parameter $q$ for FQH system defined in the previous section). A solution is found to be 
\begin{equation}
V_{m=1}^{nn-nn}(q)=\frac{\mathcal{L}_{m=1}(q^2\ell^2)}{ \left\langle n |e^{A_1}|n \right\rangle \left\langle n |e^{A_2}|n \right\rangle} \label{vq1Intra}
\end{equation}
where $\mathcal{L}_m$ is the $m$th Laguerre polynomial.
The two particle interaction matrix elements on the torus for this component of the interaction can be calculated from the real space form $V(r)=\int\text{d}\bsym{q}\, V(|\bsym{q}|)\, e^{{\iiota}\bsym{q}\cdot {(\bsym{\hat{r}}_1-\bsym{\hat{r}}_2)}}$. We find that all LL dependent information vanishes when the matrix elements are computed; so we get identical interaction matrix elements between momentum states of the torus in every LL.

The inter-LL interactions (again assuming $2p=2$) associate an energy cost whenever particles in two different LLs have a relative angular momentum $m=0$ or $1$ in the disk ie for states $\left|n,n',m=1 \right\rangle = ( (a_1^\dagger)^{n} (a_2^{\dagger})^{n'} + (a_1^\dagger)^{n'} (a_2^{\dagger})^{n}) ( b_1^\dagger - b_2^\dagger)\left | 0,0\right\rangle$ and $\left|n,n',m=0\right\rangle= ( (a_1^\dagger)^{n} (a_2^{\dagger})^{n'} - (a_1^\dagger)^{n'} (a_2^{\dagger})^{n}) \left| 0,0\right\rangle$. Here the solutions for $V(q)$ can be taken to be
\begin{align}
V^{n_1n_2-n_3n_4}_{m=1}(q) &=& \frac{\mathcal{L}_{m=1}(q^2\ell^2)(\delta_{n_1n_3}\delta_{n_2 n_4} + \delta_{n_1n_4}\delta_{n_2n_3})}{\left\langle n_1 |e^A|n_3 \right\rangle \left\langle n_2 |e^A|n_4 \right\rangle} \nonumber\\
V^{n_1n_2-n_3n_4}_{m=0}(q) &=& \frac{\mathcal{L}_{m=0}(q^2\ell^2)(\delta_{n_1n_3}\delta_{n_2 n_4} - \delta_{n_1n_4}\delta_{n_2n_3})}{\left\langle n_1 |e^A|n_3 \right\rangle \left\langle n_2 |e^A|n_4 \right\rangle} \nonumber
\end{align}
Again this leads to an interaction where torus matrix elements in the momentum space are independent of the LLs. Explicit form of the final matrix elements are given in Appendix \ref{AppToruMatEle}.

\section{Exact Eigenfunctions} \label{ExactEig}
As was argued in Sec~\ref{SecModelInt}, exact eigenfunctions of the model Hamiltonian can be constructed on the disk geometry.  In this section, we first show how to generalize these eigenfunctions to the cylinder geometry. We then discuss the sphere and torus geometries and show that similar generalization does not work in these cases. Eigenfunctions can nevertheless be written down for low-energy QP type excitations of FQH system at filling $\frac{1}{3}$ for both geometries.

\subsection{Generalizing disk eigenfunctions to cylinder} 
The unprojected CF state on the cylinder is given by
\begin{equation}
\Psi_{\nu=\frac{\nu^*}{2p\nu^*+1}}=\prod_{i< j}\rbkt{e^{\frac{2\pi {z}_i}{L}}-e^{\frac{2\pi {z}_j}{L}}}^{2p}\ \Phi_{\nu^*}
\end{equation}
where $L$ is the length along the periodic direction of cylinder, and $\Phi_{\nu^*}$ is the Slater determinant state with Landau orbitals in reduced flux $\np^*$. For the Landau gauge $\bsym{A}=-xB\hat{e}_{y}$, the single-particle state $\phi_{n,k}(\bsym{r})$ with momentum $k$ in $n$th LL, is given by
\begin{align}
\phi_{n,k}(\bsym{r})=\ &\mathcal{N}\text{exp}\rbkt{ \frac{\iiota2\pi k }{L}y} \text{exp}\sbkt{-\frac{1}{2}\rbkt{\frac{x}{\ell}-\frac{2\pi \ell}{L}k}^{2}} \times \nonumber \\
& \times H_n\rbkt{\frac{2\pi \ell}{L}k-\frac{x}{\ell}}
\end{align}
where $\mathcal{N}$ is normalization, $H_n$ is the $n$th Hermite polynomial and $k$ is the momentum quantum number which takes value in $0,1,\dots,\np-1$. 

In the spirit of the eigenfunction to our model Hamiltonian in the disk geometry, we replace the coordinates in the Jastrow factor with the guiding-center coordinates, to get
\begin{equation}
\Psi_{\nu=\frac{\nu^*}{2p\nu^*+1}}=\prod_{i< j}\rbkt{e^{\frac{2\pi \hat{Z}_i}{L}}-e^{\frac{2\pi \hat{Z}_j}{L}}}^{2p}\ \Phi_{\nu^*} \label{cylinderAnsatz}
\end{equation}
For the given gauge, the action of $\hat{Z}$ on $\phi_{n,k}(z)$ is manifested through 
\begin{equation}\label{GCOp}
e^{\frac{2\pi\hat{Z}_i}{L}} \phi_{n,k}(z)=e^{\frac{2\pi^{2}}{L^2}(2k+1)} \phi_{n,k+1}(z)
\end{equation}
Although multiplication with Jastrow factor changes the momentum state of orbitals in $\Phi_{\nu^*}$, it does not change their LL-index. Thus, it is easy to see from Eq.~\eqref{GCOp} that, just like Slater determinant $\Phi_{\nu^*}$, the ansatz $\Psi_{\nu}$ is also an eigenfunction of the kinetic energy. Kinetic energy of the state  $\Psi_{\nu}$ is identical to that of  $\Phi_{\nu^*}$, as in the case of disk geometry.

It is not straightforward to see that the ansatz defined in Eq.~\eqref{cylinderAnsatz} is a zero-energy eigenfunction of the model interaction in the cylinder geometry. For a few QPs at filling $\nu=1/3$, we numerically verified that  the eigenfunctions in Eq.~\eqref{cylinderAnsatz} match exactly with the ED eigenfunctions of the model interaction. Also, as will be shown in Sec.~\ref{NRcylinder}, the counting of low-energy states matches that the one QH spectrum, as one would expect from the ansatz in Eq.~\eqref{cylinderAnsatz}.

\subsection{Attempt to generalize disk eigenfunctions to torus} \label{IncorrectAnsatz}
Motivated by the exact eigenfunctions in the disk and cylinder geometry, we consider similar construction of the ansatz on the torus by replacing the coordinates in the Jastrow factor of the unprojected CF states,\cite{Pu2017,PeriodicLaughlin} with the guiding-center coordinates. For the torus geometry, the resulting ansatz is given by
\begin{align}
\Psi_{\nu=\frac{\nu^*}{2p\nu^*+1}}=\left[F_1(\hat{Z}_{\rm c.m.}) \right]^{2p}& {\mathcal{J}}^{2p}(\{\hat{Z}\})\  e^{-\sum_i\frac{ z^2_i+|z_i|^2 }{2\ell^2}}\ \Phi_{\nu^*} \label{Incorrectwf}
\end{align}
where
\begin{equation}
{\mathcal{J}}(\{\hat{Z}\}) = \prod_{i<j}\rtheta{1/2}{1/2}{{\hat{Z}_i-\hat{Z}_j\over L_2}}{\tau}
\end{equation} 
is the Jastrow factor of guiding-center coordinates. In section, we will write ${\mathcal{J}}^{2p}(\{\hat{Z}\})$ as $\hat{\mathcal{J}}$ for brevity. $\hat{Z}_{\rm c.m.}=\sum_{i}{\hat{Z}_i}$ is the center-of-mass of guiding-centers and $\Phi_{\nu^*}$ is the Slater determinant of Landau orbitals defined in  Eq.~\eqref{SingleParticleTorus} at flux $\np^*$ instead of $\np $. For the torus in Fig.~\ref{fig::2.1}, we have $\tau = -{L_1}/{L_2}$. $F_1(\hat{Z})$ represents the center-of-mass dependent part of the filled lowest Landau level,\cite{PeriodicLaughlin}  given by
\begin{equation}
\label{FZ}
F_1(\hat{Z}_{\rm cm})=\rtheta{\frac{\theta_2}{2\pi} + \frac{{N_e}-1}{2}}{\frac{\theta_1}{2\pi} + \frac{{N_e}-1}{2} }{\frac{\hat{Z}_{\rm cm}}{L_2}}{ \tau}
\end{equation}
where $\theta_1$ and $\theta_2$ determine  the Hilbert space representations of MTOs $t_i(L_1)$ and $t_i(L_2)$ respectively (see Sec.~\ref{SecSymTorus}). In what follows, we show that the generalization shown Eq.~\eqref{Incorrectwf}  does not yield an eigenfunction of $t_i(L_1)$ and is, therefore, not a valid eigenfunction.

To verify the boundary conditions of the ansatz, we calculate the action of the MTOs on each piece of the ansatz one-by-one. A MTO, $t_i(\bsym{a})$, can be written in terms of normal translation operators, $T_i(\bsym{a})$, in the symmetric gauge as
\begin{align}
t_i(\bsym{a}) &= e^{-\frac{\iiota}{2\ell^2}\hat{e}_z\cdot (\bsym{a}\times \bsym{r}_i)}T_i(\bsym{a}) \label{symmMTO}
\end{align}
The guiding-center coordinate, $\hat{Z}_i$, transforms under $t_i(a)$ like normal position coordinates ($z$) transform under  $T_i(\bsym{a})$. The action of $t_i(L_2)$ on various pieces of the ansatz is given as
\begin{gather}
	t_i(L_2)F_1(\hat{Z}_{\rm cm})t^{\dagger}_i(L_2)=e^{\iiota \theta_2}F_1(\hat{Z}_{\rm cm}) \label{FirstEqFail}\\
	t_i(L_2)\hat{\mathcal{J}}t^{\dagger}_i(L_2)=\hat{\mathcal{J}}
\end{gather}
The action of MTOs on functions of normal position coordinates can be calculated using Eq.~\eqref{symmMTO}, which gives us
\begin{align}
	t_i(L_2)\sbkt{e^{-\frac{\sum_i z^2_i+|z_i|^2 }{2\ell^2}}\Phi_{\nu^*}}=e^{\iiota \theta_2}\sbkt{e^{-\frac{\sum_i z^2_i+|z_i|^2 }{2\ell^2}}\Phi_{\nu^*}} \label{WrongWF1}
\end{align}
(see Appendix \ref{periodicityChecks} in  for details). By putting them together, we get
\begin{align}
t_i(L_2)\Psi_{\frac{\nu^*}{2p\nu^*+1}}&=e^{i\alpha\theta_2}\Psi_{\frac{\nu^*}{2p\nu^*+1}}\label{pb1}
\end{align}
where $\alpha=(2p+1)$. This implies that $\Psi$ is an eigenfunction of $t_i(L_2)$. Now we consider the action of $t_i(L_1)$ on the ansatz $\Psi$.  It is easy to check that
\begin{gather}
t_i(L_1)F_1(\hat{Z}_{\rm cm})t^{\dagger}_i(L_1)=e^{\iiota \theta_1}e^{-\iiota \pi \tau}e^{\frac{ \iiota 2\pi \hat{Z}_{\rm cm}}{L_2}} F_1(\hat{Z}_{\rm cm}) \label{t1_1}\\
t_i(L_1)\hat{\mathcal{J}}t^{\dagger}_i(L_1)=e^{-\iiota \pi(N_e-1) \tau}e^{-\frac{ \iiota 2\pi \hat{Z}_{\rm cm}}{L_2}}e^{\frac{ \iiota 2\pi N_e\hat{Z}_i}{L_2}}\hat{\mathcal{J}} \label{t1_2}
\end{gather}
and
\begin{align} 
t_i(L_1)\sbkt{e^{-\frac{\sum_i z^2_i+|z_i|^2 }{2\ell^2}}\Phi_{\nu^*}}&=\nonumber \\
e^{\iiota \theta_1}e^{\iiota 2\pi pN_e \tau }e^{-\frac{ \iiota 4\pi p N_e{\hat{Z}}_i}{L_2}}&\sbkt{e^{-\frac{\sum_i z^2_i+|z_i|^2 }{2\ell^2}}\Phi_{\nu^*}} \label{lastEqFail}
\end{align}
By combining these results together, the action of $t_i(L_1)$ on the ansatz is found to be
\begin{align}
t_i(L_1)\Psi_{\frac{\nu^*}{2p\nu^*+1}}&=e^{i\alpha\theta_1}e^{{\iiota 4\pi pN_e\over L_2}(\hat{Z}_i-z_i)}\Psi_{\frac{\nu^*}{2p\nu^*+1}}\label{pb2}
\end{align}
Since the action of $t_i(L_1)$ on ansatz results in a factor which contains the guiding-center operator $\hat{Z}_i$, the ansatz is not an eigenfunction of $t_i(L_1)$ and hence not a valid state on torus. Using Eq.~\eqref{t1_1}-\eqref{lastEqFail} it can be verified that, similar factor will arise even if we use $F_1({Z}_{\rm c.m.})$ with normal center-of-mass coordinate instead, in the ansatz. In summary, this implies that the ansatz obtained by the straightforward generalization [Eq.~\eqref{Incorrectwf}] of Eq.~\eqref{diskansatz} valid on disk and Eq.~\eqref{cylinderAnsatz} valid on cylinder, does not yield an eigenfunction on torus.

\subsection{Exact Eigenfunction for QPs of $\nu= \frac{1}{3}$}\label{correctAnsatz}
As shown in the previous section the disk ansatz Eq.~\eqref{diskansatz} does not generalize to the torus. It is also not possible to generalize it to spherical geometry as the form of the guiding-center coordinate is not known for the sphere. 
In this section, we show that for a subset of states namely quasiparticles of the $1/3$ state, the ansatz can be written in a simplified form, which generalizes to sphere and torus geometries. For the disk and the spherical geometry we could verify that the results from the ED of the Hamiltonian (Eq.~\eqref{DiscHamiltonian}) match with the form of the eigenfunction presented here. Finally, we show that a similar generalization leads to a valid state on torus, as it satisfies the periodic boundary conditions.

\subsubsection{A different point of view of disk eigenfunctions} \label{altAnsatzDisKandSphere}
The ansatz in Eq.~\eqref{diskansatz} for $N$ QPs of $1/3$ can be written as
\begin{equation}
\Psi^{ N \rm -QPs}_{1/3}=\mathcal{J}^{2}(\{\hat{Z} \}) \times \Phi^{ N \rm-QPs}_{1} \label{1b3discAnsatz}
\end{equation}
where $\mathcal{J}(\{\hat{Z}\})=\prod_{i<j} (\hat{Z}_i - \hat{Z}_j)$ is the Jastrow factor of guiding-center coordinates, and Slater determinant $\Phi^{ N \rm-QPs}_{1}$ contains $N$ particles in LL1 with fully occupied LLL. The Landau orbitals at angular momentum $m$ in LLL and LL1 are represented by $F_{0,m}$ and $F_{1,m}$ respectively (We apologize for using the same symbols as the center-of-mass part, but they are different).  We show in Appendix \ref{Ansatz1by3QPDisc} that the wave function $\Psi^{ N \rm-QPs}_{1/3}$ can be rewritten as
\begin{equation}
\Psi^{ N\rm-QPs}_{1/3}= \hat\Phi^{ N\rm-QPs}_{1} \times {\mathcal{J}}^{2}(\{z\}) \label{AlterSolDisk}
\end{equation}
The operator $\hat\Phi^{ N\rm-QPs}_{1}$ which acts on $\mathcal{J}^{2}(\{z\})$ can be constructed by replacing any LL1 orbitals $F_{1,m}$, inside $\Phi^{ N\rm-QPs}_{1}$ with $\hat{G}_{1,m}=F_{1,m} -  \hat{F}_{1,m}$.  The operator $\hat{F}_{1,m}$ is  defined such that, for any momentum $k$
\begin{equation}
\hat{F}_{1,m}  F_{0,k}= \mathcal{P}_{\rm LLL} \sbkt{ F_{1,m}  F_{0,k}}
\end{equation}
where $ \mathcal{P}_{\rm LLL}$  is the LLL projection operator. We get operator $\hat{F}_{1,m}$ by replacing all $\bar{z}$ in  ${F}_{1,m}(z,\bar{z})$ with $2\ell^2\partial_z$, after all the $z$'s are moved to the left.\cite{Jain07} It should be noted that the derivatives $\partial_z$'s do not act on the exponential factor.  The operator $\hat{G}_{1,m}$ can be better understood as
\begin{gather}
\hat{G}_{1,m} F_{0,k}  = \sbkt{ F_{1,m} -  \hat{F}_{1,m} }  F_{0,k}\nonumber\\ 
= F_{1,m} F_{0,k}- \mathcal{P}_{\rm LLL} \sbkt{ F_{1,m} F_{0,k}}    = \mathcal{P}_{\rm LL1} \sbkt{ F_{1,m} F_{0,k}} \label{LL1proj}
\end{gather} 
In summary, the operator $\hat\Phi^{ N\rm-QPs}_{1}$ in the last expression of Eq.~\eqref{AlterSolDisk} is given by
\begin{multline}\label{deterPhiAlpha}
\hat\Phi^{ N\rm-QPs}_{1}=\\ \left\vert \begin{matrix}
F_{0,0}(z_1)  & F_{0,0}(z_2) & \dots & F_{0,0}(z_{N_e}) \\
F_{0,1}(z_1)  & F_{0,1}(z_2) & \dots & F_{0,1}(z_{N_e}) \\
\vdots & \vdots & \vdots & \vdots \\ 
F_{0,\np^*-1}(z_1)  & F_{0,\np^*-1}(z_2) & \dots & F_{0,\np^*-1}(z_{N_e}) \\
\hat{G}_{1,m_{1}}(z_1,\partial_{z_1})  & \hat{G}_{1,m_{1}}(z_2,\partial_{z_1}) & \dots & \hat{G}_{1,m_{1}}(z_{N_e},\partial_{z_{N_e}})\\
\vdots & \vdots & \ddots & \vdots \\ 
\hat{G}_{1,m_{N}}(z_1,\partial_{z_1})  & \hat{G}_{1,m_{N}}(z_2,\partial_{z_1}) & \dots & \hat{G}_{1,m_{N}}(z_{N_e},\partial_{z_{N_e}})
\end{matrix}
\right\vert
\end{multline}

Although, it is not as easy to see that the alternate form of $\Psi^{\alpha}$ defined in Eq.~\eqref{AlterSolDisk} are zero-energy eigenfunctions of the model interaction, these can be numerically evaluated for small systems. Upon comparison with exact diagonalization spectrum of the model Hamiltonian given in Eq.~\eqref{DiscHamiltonian} on the disk, we could explicitly verify that these are the right eigenfunctions.

More importantly, the expression in  Eq.~\eqref{AlterSolDisk} can be written for sphere as well:
\begin{equation}
	\Psi^{\rm N-QPs}_{1/3} = \hat{\Phi}^{\rm N-QPs}_{1/3}\rbkt{Y_{Q^*1m}\rightarrow Y_{Q^*1m}-\hat{Y}^{q}_{Q^*1m}} \mathcal{J}^{2} \label{SphereAltAnsatz}
\end{equation}
where monopole harmonics $Y_{Qnm}$ represent single-particle Landau orbitals with angular momentum $m$ in $n$th LL, in flux given by $2Q$. For any LLL state $Y_{q0k}$, the operator $\hat{Y}^{q}_{Q^*1m}$ is defined using
\begin{equation}
\hat{Y}^{q}_{Q1m}  Y_{q0k}= \mathcal{P}_{\rm LLL} \sbkt{ Y_{Q1m} Y_{q0k}}
\end{equation}

Again, we could explicitly compare this with ED eigenfunctions for small systems and verify that the ansatz in Eq.~\eqref{SphereAltAnsatz} gives correct eigenfunctions. In summary, Eq.\eqref{AlterSolDisk} and \eqref{SphereAltAnsatz} describe QPs of the $1/3$ state on disk and sphere geometries.

\subsubsection{Ansatz for torus geometry}
Motivated by the applicability of ansatz in  Eq.\eqref{1b3discAnsatz} to the spherical geometry, in this section we ask whether it also gives a valid state on torus. For $N$ QPs on $\nu=1/3$ with $N_e$ particles in $\np$ flux quanta, the analog of Eq.\eqref{1b3discAnsatz} in torus geometry is
\begin{multline}
\Psi^{\rm N-QPs}_{1/3}= e^{-\frac{\sum_i^{{N_e}} z^2_i+|z_i|^2  }{2\ell^2}}\ \sbkt{F_1(Z_{\rm c.m.})}^2  \times \\
  \times \hat{\mathcal{D}}^{ N\rm-QPs}_{1}\rbkt{\{f^k_i,\hat{g}^q_j\}} \mathcal{J}^{2}(\{z\}) \label{correctWF}
\end{multline}
where $Z_{\rm c.m.}=\sum_i z_i$ and $\mathcal{J}(\{z\}) $ is the Jastrow factor, given by
\begin{equation}
\mathcal{J}(\{z\}) = \prod_{i<j}^{N_e} \rtheta{1/2}{1/2}{\frac{z_i-z_j}{L_2}}{\tau}
\end{equation}
All states inside the determinant $\hat{\mathcal{D}}^{ N\rm-QPs}\rbkt{\{f^k_i,\hat{g}^q_j\}}$  see reduced flux $\np^{*} =\np - 2N_e$ and the corresponding magnetic length is given by $\ell^{*}$, defined as $({\ell^{*}})^{2}=\ell^2{\np}/\np^{*}$.  For $N$ QPs in 2LL, the operator  $\hat{\mathcal{D}}^{ N\rm-QPs}_{1}\rbkt{\{f^k_i,\hat{g}^q_j\}}$ is defined as
\begin{equation}\label{deter}
= \left\vert \begin{matrix}
f^{0}_{0}(z_1) & f^{0}_{0}(z_2) & \dots & f^{0}_{0}(z_{N_e}) \\
f^{1}_{0}(z_1) & f^{1}_{0}(z_2) & \dots & f^{1}_{0}(z_{N_e}) \\
\vdots & \vdots & \vdots & \vdots \\ 
f^{{\np^*-1}}_{0}(z_1) & f^{{\np^*-1}}_{0}(z_2) & \dots & f^{{\np^*-1}}_{0}(z_{N_e}) \\

\hat{g}^{q_1}_{1}(z_1,\bar{z}_1) & \hat{g}^{q_1}_{1}(z_2,\bar{z}_2) & \dots & \hat{g}^{q_1}_{1}(z_{N_e},\bar{z}_{N_e}) \\
\vdots & \vdots & \ddots & \vdots \\ 
\hat{g}^{q_{N}}_{1}(z_1,\bar{z}_1) & \hat{g}^{q_{N}}_{1}(z_2,\bar{z}_2) & \dots & \hat{g}^{q_{N}}_{1}(z_{N_e},\bar{z}_{N_e}) \\
\end{matrix}
\right\vert
\end{equation}
where $f_0^{k}(z)$ and $f^{k}_1(z,\bar{z})$ represent the single particle wave functions in LLL and LL1 respectively, and are defined as \cite{Pu2017}
\begin{gather}
\quad f_0^{k}(z) = \rtheta{\frac{k}{\np^*} +\frac{\theta_2}{2\pi \np^*}}{\frac{\theta_1}{2\pi}}{\frac{\np^* z}{L_2} }{\np^*\tau} \nonumber\\
f^{k}_1(z,\bar{z})= \sqrt{2}{\ell^{*}}\sbkt{\frac{\bar{z}+z}{2(\ell^*)^{2}} -\partial_z} f^{k}_0(z) \label{singleParticleOrbitals}
\end{gather}
The operator $\hat{g}^{q}_1(z,\bar{z})=f^{q}_1(z,\bar{z})-\hat{f}^{q}_1(z,\bar{z})$ is analogous to the $\hat{G}$ operator defined in Eq.\eqref{LL1proj}. The operator $\hat{f}^{q}_1(z,\bar{z})$ is defined as
\begin{equation}
\hat{f}^{q}_1(z,\bar{z}) = \sqrt{2}\ell^* \sbkt{-2\nu \partial_z f^{q}_0(z) + (1-2\nu)f^{q}_0(z) \partial_z}
\end{equation}
which implies that $\hat{g}^{q}_1(z,\bar{z})$ is given by
\begin{equation} \label{ghat}
\hat{g}^{q}_1(z,\bar{z})=  \frac{\sqrt{2}\np^*\ell^*}{\np} \sbkt{ \frac{\bar{z}+z}{2\ell^{2}}f^{q}_0(z) - \partial_z f^{q}_0(z) - f^{q}_0(z) \partial_z} \nonumber
\end{equation}
such that $\hat{g}^{q}_1 f_0^{k}=\mathcal{P}_{LL1} \sbkt{f^{q}_1 f_0^{k}}$.

Now we show that the state defined in Eq.~\eqref{correctWF} satisfies the periodic boundary conditions. We calculate the action of MTOs $t_i(L_1)$ and $t_i(L_2)$ on different parts of the ansatz (see Appendix \ref{periodicityChecks} for details). First, the action on the exponential factor is given by 
\begin{gather}
t_i(L_2){e^{-\frac{\sum_i z^2_i+|z_i|^2 }{2\ell^2}}}= {e^{-\frac{\sum_i z^2_i+|z_i|^2 }{2\ell^2}}} \label{tl11}\\
t_i(L_1){e^{-\frac{\sum_i z^2_i+|z_i|^2 }{2\ell^2}}}= e^{\iiota \pi \np \rbkt{\tau-\frac{2z_i}{L_2}} }{e^{-\frac{\sum_i z^2_i+|z_i|^2 }{2\ell^2}}}\label{tl21}
\end{gather}
Their action on $F_1(Z_{\rm cm})$ is given by
\begin{gather}
t_i(L_2)F_1(Z_{\rm cm})t^{\dagger}_i(L_2)=e^{\iiota \theta_2}F_1(Z_{\rm cm}) \label{tl12}\\
t_i(L_1)F_1(Z_{\rm cm})t^{\dagger}_i(L_1)=e^{\iiota \theta_1}e^{-\iiota \pi \tau}e^{\frac{ \iiota 2\pi Z_{\rm cm}}{L_2}} F_1(Z_{\rm cm}) \label{tl22}
\end{gather}
Since the Slater determinant $\hat{\mathcal{D}}^{ N\rm-QPs}_{1}\rbkt{\{f^k_i,\hat{g}^q_j\}}$ contains operators, the action of MTOs is rather calculated on the combined piece \ie  $\hat{\mathcal{D}}^{ N\rm-QPs}_{1}\rbkt{\{f^k_i,\hat{g}^q_j\}}\mathcal{J}^2$. 

In the expansion of the Slater determinant, the $i$th particle can either be in LLL, in which case it will be represented by some state $f_0^{k_j}(z_i)$, or it can be in second LL, where it will be an operator $\hat{g}_1^{q_j}(z_i)$. The operators $\hat{g}_1^{q_j}(z_i)$'s commute with each other for different particles.  Since, the single particle MTO, $t_i(a)$, only affects the $i$th particle, we only need to check the action on $f_0^{k_j}(z_i)\mathcal{J}^2$ and $\hat{g}_1^{q_j}(z_i)\mathcal{J}^2$, which are given by 
\begin{align}
t_i(L_2)f_0^{k_j}(z_i)\mathcal{J}^2=&\ e^{\iiota \theta_2}  f_0^{k_j}(z_i)\mathcal{J}^2 \label{tl13}\\
t_i(L_2)\hat{g}_1^{q_j}(z_i)\mathcal{J}^2 =&\ e^{\iiota \theta_2} \hat{g}_1^{q_j}(z_i)\mathcal{J}^2 \label{tl14}
\end{align}
Using Eqs.~\eqref{tl11}, \eqref{tl12}, \eqref{tl13} and \eqref{tl14}, we get 
\begin{align}
t_i(L_2)\Psi_{\frac{\nu^*}{2p\nu^*+1}}&=e^{i3\theta_2}\Psi_{\frac{\nu^*}{2p\nu^*+1}}
\end{align}
Similarly, the action of $t_i(L_1)$ is given by
\begin{align}
t_i(L_1)&f_0^{k_j}(z_i)\mathcal{J}^2=\  e^{\iiota \theta_1}e^{\iiota\pi(2-\np)\tau}  e^{ \frac{\iiota 2\pi}{L_2} (\np z_i-2Z_{\rm cm})} \times \nonumber\\
&\quad\quad \times f_0^{k_j}(z_i)\mathcal{J}^2 \label{tl23}\\
t_i(L_1)&\hat{g}_1^{q_j}(z_i)\mathcal{J}^2=\ e^{\iiota \theta_1} e^{\iiota\pi(2-\np)\tau}  e^{ \frac{\iiota 2\pi}{L_2} (\np z_i-2Z_{\rm cm})} \times \nonumber\\
 &\quad \quad  \times   \sbkt{ \hat{g}_1^{q_j}(z_i)-\frac{\iiota 4\pi A (N_e-1)}{L_2}f_0^{q_j}(z_i)} \mathcal{J}^2 \label{ghatProblem}
\end{align}
where $A={\sqrt{2}\np^*\ell^*}/{\np}$. There are two terms which could cause the boundary conditions to not be satisfied: First, in Eq.~\eqref{ghatProblem}, we get an addition term $-{\iiota 4\pi A(N_e-1)}f_0^q(z_i)/{L_2}$ along with $\hat{g}_1^q(z_i)$ inside the square bracket. Secondly, if there are more that one QPs in the system,  $\hat{g}_1^q(z_j)$'s for other QPs will act on the factor $e^{ -\frac{\iiota 4\pi Z_{\rm cm}}{L_2}}$ and produce further terms. These are equivalent to replacing $\hat{g}_1^{q_j}(z_i)$ for the $i$th particle with $\hat{g}_1^{q_j}(z_i) + a f_0^{q_j}(z_i)$ and $\hat{g}_1^{q_j}(z_k)$'s for $k\neq i$ with $\hat{g}_1^{q_j}(z_k) + b  f_0^{q_j}(z_k)$ in the Slater determinant $\hat{\mathcal{D}}\rbkt{\{f^k_i,\hat{g}^q_j\}}$, where $a$ and $b$ are constants for a given problem. Since all the LLL states $f_0^{q_j}$'s are filled, the terms of kind $ a f_0^{q_j}(z_i)$ and $b f_0^{q_j}(z_k)$ can be removed without affecting the determinant $\hat{\mathcal{D}}^{ N\rm-QPs}_{1}$. Putting everything together from Eqs.~\eqref{tl21}, \eqref{tl22}, \eqref{tl23} and \eqref{ghatProblem}, we get
\begin{align}
t_i(L_1)\Psi_{\frac{\nu^*}{2p\nu^*+1}}&=e^{i3\theta_1}\Psi_{\frac{\nu^*}{2p\nu^*+1}}
\end{align}
which means that it satisfies the torus boundary conditions. Note that we have shown that the wave function in Eq.~\eqref{correctWF} is a valid torus state. The state is an eigenfunction of kinetic energy $H_{\rm KE}$, 
however we have not been able to check that the state is a zero energy eigenfunction of the model interaction on torus. Hence this is a conjecture supported by the validity of ansatz expression in the disk and more importantly in the spherical geometry.

\section{Numerical Results} \label{NumRes}
In Ref.~\onlinecite{Anand20}, we explored the spectra of the model Hamiltonian in the spherical geometry and explicitly demonstrated the correspondence between spectrum of the model Hamiltonian and the IQH spectrum. In this work, we compute the same in torus and cylinder geometries.   

First, we present and discuss the features low-energy spectra for the model interaction on the torus geometry. In the results shown below, we consider different Hall liquids are labeled by $(\np,N_e)$ configurations. The eigenfunctions are labeled using $K_2$ and $\tilde{K}_1$ which are quantum numbers associated to MTOs $t_{\rm c.m.}\rbkt{\bsym{L}_2/\np}$ and ${\tilde{t}}_i (p \bsym{L}_1)$ (Sec.~\ref{SecSymTorus}). The following results are for a square torus where $|L_1|=|L_2|$ and $L_\Delta=0$, which implies  $\tau=\iiota$.

\subsection{Spectra on the torus geometry}
The CF wave functions\cite{Jain89} describe FQHE systems at Jain sequence filling fraction $\nu=n/(2pn+1)$ by mapping the interacting system of $N_e$ electrons, in flux $\np$ (in the units of flux quanta $\phi_0=2\pi \hbar/e$), to a non-interacting system of CFs in a reduced flux given by $\np^{*}=\np-2pN_e$. While in Ref.~\onlinecite{Anand20}, we showed that in spherical geometry there is a one-to-one correspondence between IQH and model Hamiltonian spectra, in this section we will show that a similar map exists for torus geometry as well, but instead there is a one-to-$(2pn+1)$ mapping present in the IQH and FQH spectra for the torus geometry.

From the spectra shown presented in the subsequent section, we infer the following key results. For a system with $N_e=pN$ particles in $\np=qN$ flux quanta, where $N=\text{gcd}(\np,N_e)$, the low-energy spectra of the model Hamiltonian, in a given $(\tilde{K}_1,K_2)$-sector, is identical to the $(\tilde{K}_1,K_2^{I})$-sector spectra of a non-interacting system in a reduced flux $\np^*$, where $K_2$ and $K^I_2$ are related by
\begin{gather}
 K_2=K_2^{I} + rN\quad \quad\text{}r=0,1,\dots,q-1 \label{4.1}
\end{gather}
and $K_2\in [0,\np)$;  $K_2^{I}\in [0,\np^{*})$ is the quantum number corresponding to $t_{\rm c.m.}(\bsym{L}_2/\np^*)$ for the IQHE system. We show this equivalence between spectra in several cases below. 

\begin{figure}[h!]
	\includegraphics[width=0.80\columnwidth]{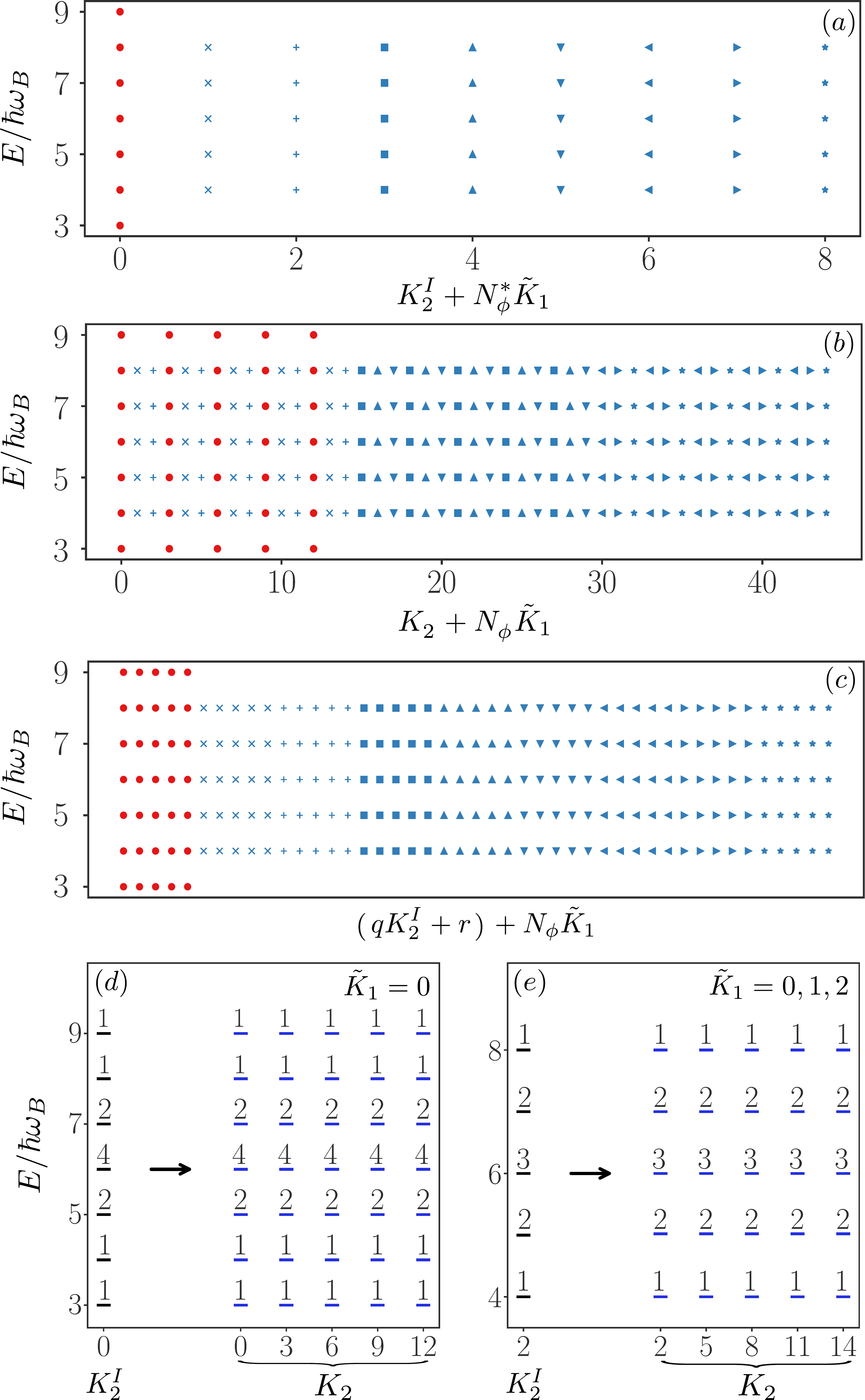}
	\caption{
  $(a)$ Shows the spectrum for a non-interacting system (IQH) at integer filling factor $\nu^{*}=2$, with $N_e=6$ particles in flux $\np^{*}=3$. Pair of quantum numbers $(\tilde{K}_1,K_2^{I})$ label each state along the $x$-axis. Spectrum in each $(\tilde{K}_1,K_2^{I})$-sector is represented by a different marker. In addition, different colors  are assigned for unique degeneracy patterns along the $y$-axis. Spectra with red and blue markers have degeneracy pattern of $(1,1,2,4,2,1,1)$ and  $(1,2,3,2,1)$, respectively. The state with energy $E/\hbar\omega_B=3$ represents the incompressible ground state at integer filling factor $\nu^{*}=2$  whereas states with $E/\hbar\omega_B=4$ contain a single neutral excitation.  $(b)$ Shows the spectra for ZIE eigenspace of the model interaction (FQH) for system with $N_e=6$ particles in flux $\np=15$ at $\nu=p/q=\frac{2}{5}$. The states are labeled with quantum numbers $(\tilde{K}_1,K_2)$ along the $x$-axis and are represented by the same markers used for the $(\tilde{K}_1,K_2^{I})$-sector of IQH spectrum if $K_2$ and $K_2^{I}$ satisfy Eq.~\eqref{4.1}. We see that, apart from the fivefold topological degeneracy, the spectrum is identical to the IQH spectra. $(c)$ Shows the same spectrum using a different arrangement of $(\tilde{K}_1,K_2)$ along the $x$-axis where the $1$-to-$5$ correspondence with IQH spectrum in panel $(a)$ is more evident. Here definitions of $r,\ q$ are same as in Eq.~\eqref{4.1}. $(d)$, $(e)$ Show  maps of IQH spectrum (black) in two different $(\tilde{K}_1,K_2^{I})$-sectors, representing two unique degeneracy patterns present in the full spectra. This is juxtaposed with the spectrum of the FQH system at corresponding $(\tilde{K}_1,K_2)$-sectors satisfying Eq.~\eqref{4.1}.}\label{fig3}
\end{figure}

\paragraph{Incompressible state at $\nu=2/5$:} In Fig.~\ref{fig3}$(a)$ and  Fig.~\ref{fig3}$(b)$, we show the spectra of non-interacting system at $\nu^{*}=2$ and the low-energy spectra of model interaction at $\nu=2/5$, respectively. FQH and IQH systems have $N_e=6$ particles in flux $\np=15$ and $\np^*=3$, respectively. Each marker represents an eigenfunction (or eigenfunctions, when degenerate) with its energy shown along the $y$-axis. The sectors which these eigenfunction (or eigenfunctions) belong to are represented by a unique combination of quantum numbers $(\tilde{K}_1,K^I_2)$ and $(\tilde{K}_1,K_2)$, along the $x$-axis, for IQH and FQH system respectively. For visible distinction, states in $(\tilde{K}_1,K^I_2)$-sectors are color-coded according to their degeneracy pattern along the energy axis. 

For instance, the spectrum for sector $(\tilde{K}_1,K^I_2)=(0,0)$ of IQH has a degeneracy pattern of $(1,1,2,4,2,1,1)$ at energies $E/\hbar\omega_B=(3,4,5,6,7,8,9)$. We have shown all sectors with this pattern in red color.  The same degeneracy pattern is seen in the model Hamiltonian spectrum, in $(\tilde{K}_1=0,K_2)$-sectors where the $K_2$-values are given by $0,3,6,9,12$, as expected from Eq.~\eqref{4.1}. There can be more than one $(\tilde{K}_1,K^I_2)$-sectors with the same degeneracy pattern. Equation~\eqref{4.1} suggests that FQH states in each $(\tilde{K}_1,K_2)$-sector can be uniquely labeled by $(\tilde{K}_1,K^I_2,r)$. All FQH states labeled with same $(\tilde{K}_1,K^I_2)$ have the same spectrum and $r$ takes values in $0,1,\dots,q-1$. In Fig.~\ref{fig3}$(c)$, the choice of $x$-axis ensures that sectors with the same $\tilde{K}_1$ and $K^I_2$ appear together, allowing us to clearly see the $1$-to-$5$ correspondence.

Figures ~\ref{fig3}$(c)$ and ~\ref{fig3}$(d)$ show the map between  $K^I_2$ and $q=5$ corresponding $K_2$ sectors for two different representative cases. Spectra for IQH states in a given  $(\tilde{K}_1,K^I_2)$-sector is shown in black whereas spectra of  FQH in different  $(\tilde{K}_1,K_2)$-sectors are shown in blue. The $K^I_2$ quantum number for IQH states and $q=5$ different $K_2$ quantum numbers for FQH states, satisfying Eq.~\ref{4.1}, are labeled along the $x$-axis.

Figures \ref{fig3}$(c)$ and \ref{fig3}$(d)$ show the spectra at one $(\tilde{K}_1,K_2^{I})$-sector of IQH together with the spectrum at sector $(\tilde{K}_1,K_2^{I} + rN)$ for $r=0,1,\dots,q-1$. We note that these sectors have identical spectrum validating the relation in Eq.~\eqref{4.1}. 

\begin{figure}[h]
	\includegraphics[width=\columnwidth]{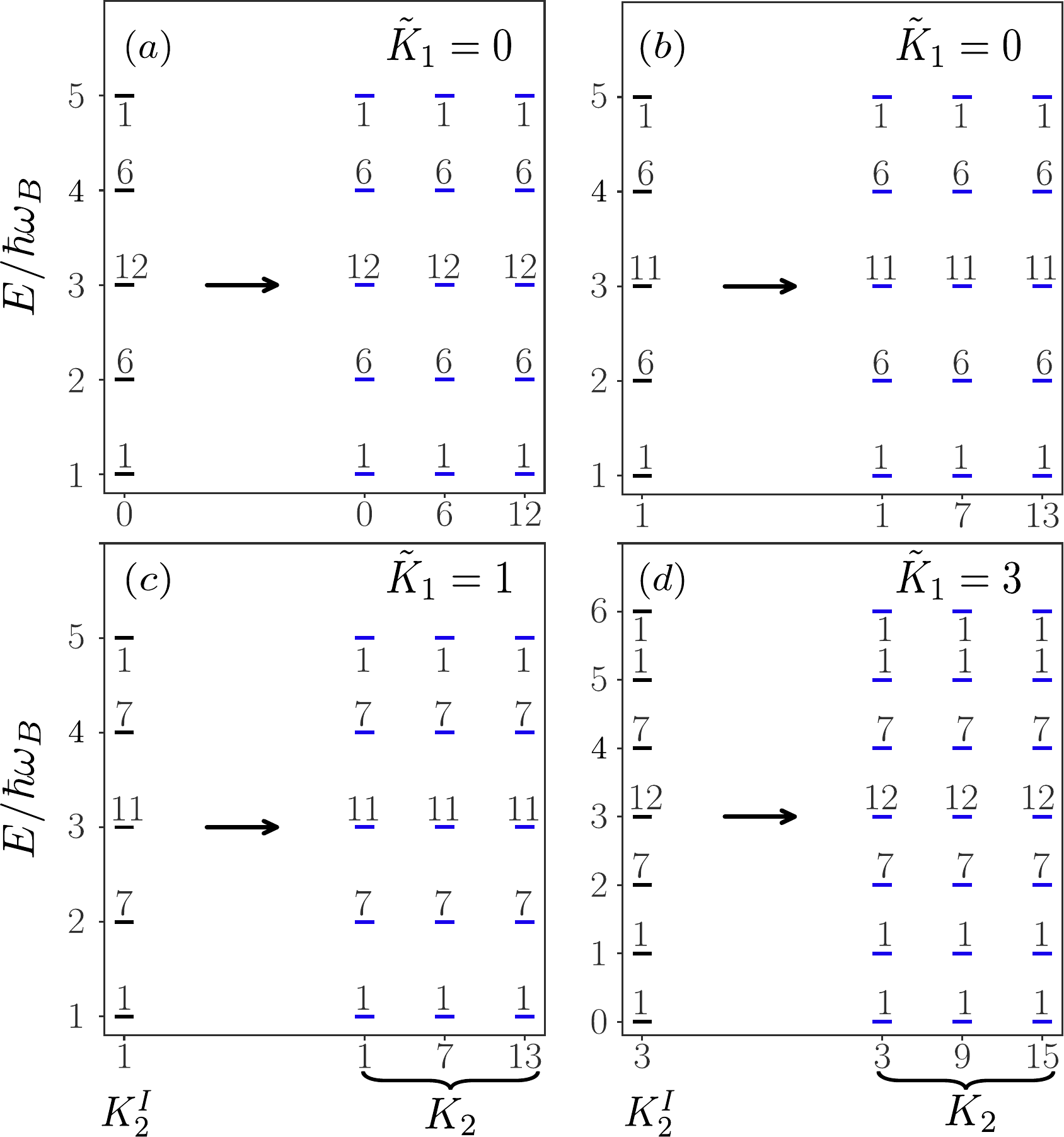}
	\caption{ Plot showing the IQH-to-FQH mapping in low-energy spectra for IQH system (black) with $(\np^{*},N_e)=(6,6)$, at integer filling factor $\nu^{*}=1$, to the corresponding FQH spectra (orange) with $(\np,N_e)=(18,6)$ at $\nu=1/3$. The full IQH spectra consists of four different degeneracy patterns, and each panel shows their mapping to the corresponding $K_2$-sectors for FQH system: $(a)$ in $\tilde{K}_1=0$ sector, IQH spectra for ${K}_2^{I}=0$ sector maps to those with ${K}_2=0,6,12$ sectors in FQH system. Panels $(b)$ and $(c)$ show similar maps for other unique spectra in given sectors. Panel $(d)$ shows map of spectra which contains  zero-energy state corresponding to the incompressible ground state of $\nu=1/3$. Full spectrum is shown in Fig.~\ref{figA1} of Appendix \ref{AdditionalFigures}.
	}
	\label{fig4}
\end{figure}

\paragraph{Incompressible state at $\nu=1/3$}
Full spectrum for $1/3$ state for system with $N_e=6$ and $\np=18$ is given in Appendix \ref{AdditionalFigures} (Fig.~\ref{figA1}). Fig.~\ref{fig4} shows the mapping between $(\tilde{K}_1,K^I_2)$-sectors in IQH spectra and $q=3$ different $(\tilde{K}_1,K_2)$-sectors in the FQH for $1/3$ state. There are four panels, one for each $(\tilde{K}_1,K^I_2)$-sector of IQH spectra representing a unique degeneracy-pattern present in the full IQH spectra (Fig.~\ref{figA1} in  Appendix \ref{AdditionalFigures}). All of these map to $q=3$ different $K_2$-sectors in FQH, which follow Eq.~\ref{4.1}. Similar results for the incompressible state at $\nu=\frac{3}{7}$ and
QP/QH excitations of $\nu=\frac{1}{3}, \frac{2}{5}$ are given in
Fig.~\ref{figA1} and Fig.~\ref{figA2} of Appendix \ref{AdditionalFigures}, respectively.

\paragraph{For charged excitations of $\nu=1/3,\ 2/5$}
The $1$-to-$q$ mapping between corresponding sectors in the IQH and FQH spectra also holds for quasi particles (QPs) and quasi-holes (QHs) of filling fractions $\nu=1/3$ and $2/5$. Details of the systems and their spectra are given in Appendix \ref{AdditionalFigures}.

\subsection{Spectra on cylinder geometry} \label{NRcylinder}
In cylinder geometry, the single particle state is labeled by linear momentum  $k$ due to translation invariance along circumference of size $L$. Unlike the torus, the cylinder does not have any non-trivial many-body translation symmetries, hence the spectra of model interaction are only indexed by $K_{\rm total}=\sum_i k_i$ where $k_i\in[0,\np)$ and $\np$ is the maximum number of orbitals in each LL. The minimum and maximum values of $k_i$ are $k_{\rm min}=0$ and $k_{\rm max}=\np-1$.

\begin{figure}[h]
	\includegraphics[width=\columnwidth]{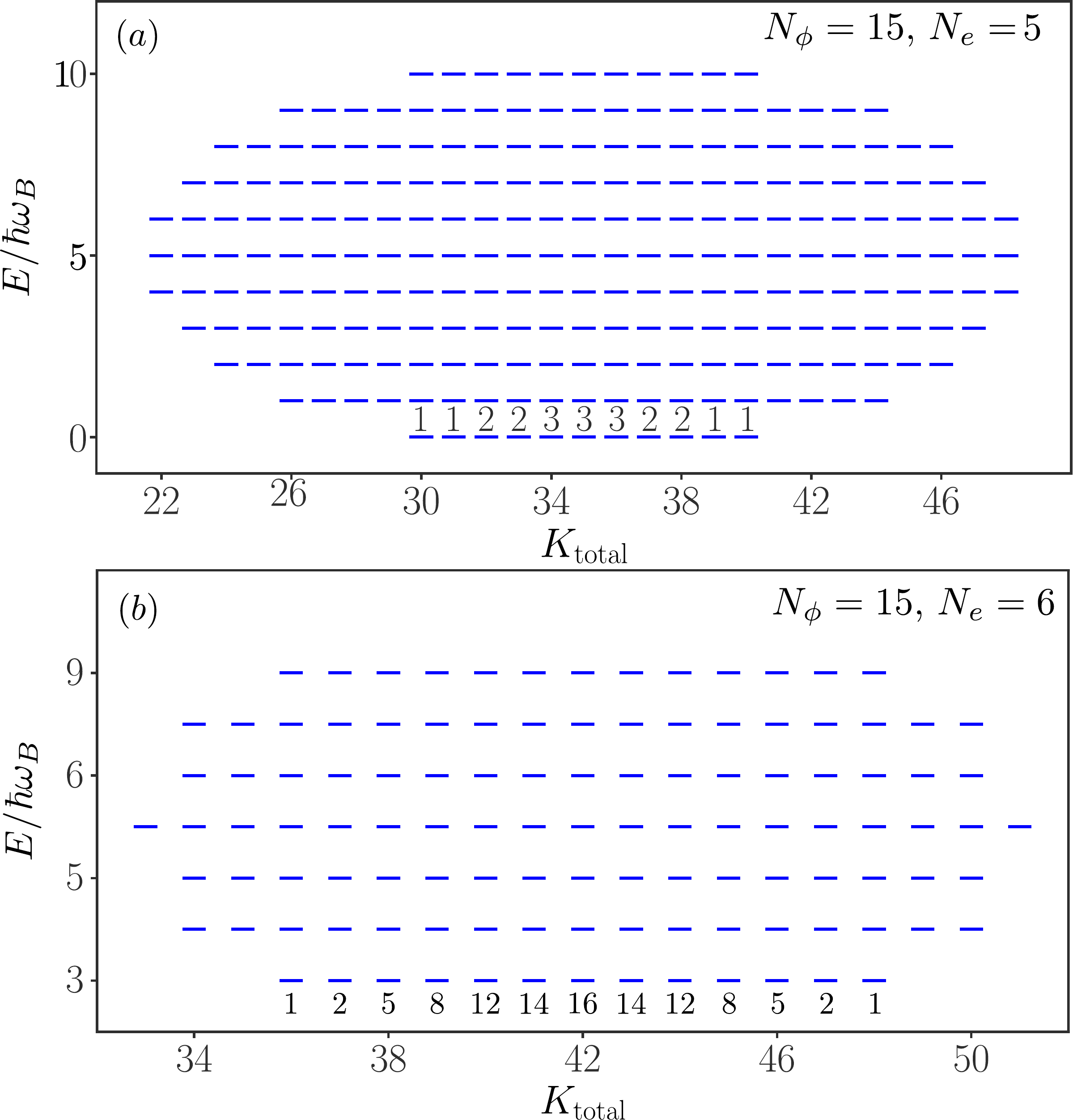}
	\caption{
		Spectra for the model Hamiltonian in cylinder geometry, for FQH systems at filling $\nu=1/3$ $(a)$ and $2/5$  $(b)$. 
		The $x$-axis labels the $K_{\rm total}$ quantum number, and energy $E/\hbar\omega_B$ is along the $y$-axis. Lowest $3$ LLs are used in these calculations.
		Panel $(a)$ shows the spectrum for system with $\np=15$ and $N_e=5$. The numbers above states with $E/\hbar\omega_B=0$ represent their degeneracy. The state at $K_{\rm total}=30$ corresponds to the incompressible ground state at filling $\nu=1/3$ and other states are its QH/edge and center-of-mass excitations. States at higher energy correspond to neutral excitations of $1/3$. Similarly, $(b)$ shows the  spectrum for system with $\np=15$ and $N_e=6$. Here the state at $K_{\rm total}=36$ with $E/\hbar\omega_B=3$ is the ground state for $\nu=2/5$.
	}
	\label{CylSpectra}
\end{figure}

In the top panel of Fig.~\ref{CylSpectra}, we show the spectrum for system with $N_e=5$ in flux $\np=15$, at filling fraction $\nu=1/3$. The eigenfunction with zero energy at  $K_{\rm total}=30$ corresponds to the incompressible ground state of $1/3$. At the same energy, the eigenfunctions at higher $K_{\rm total}$ are the quasi-hole/edge excitations as well as center-of-mass excitations where the numbers represent the degeneracy at a given $K_{\rm total}$-value. The counting at small momenta match that of the edge excitation of $\nu=1/3$. At large momenta the counting deviates due to finite system size. The states in the higher energy bands are the neutral excitations of $\nu=1/3$. Similarly, the lower panel shows the spectra of state at $\nu=2/5$ filling, with $N_e=6$ at flux $\np=15$. Here the incompressible ground state with energy $E/\hbar\omega_B=3$ is at $K_{\rm total}=36$. Again, the same energy band shows QH/edge and center-of mass excitations at larger $K_{\rm total}$ values, and the higher energy band hosts neutral excitations of the $2/5$ FQH state.

\section{Conclusion} \label{conclusion}

In this work, we extend the ideas presented in Ref \onlinecite{Anand20} to torus and cylinder geometry. The model Hamiltonian (Eq.~\eqref{diskModelInter}) introduced there was written in the disk geometry and studied in the disk and spherical geometry. 
The Hamiltonian has some appealing features - a single Hamiltonian produces incompressible states at all Jain filling fractions of the form $n/(2pn+1)$ and allows exact eigenfunctions for the incompressible states, quasihole states, quasiparticle and neutral excitations. The spectrum of the system at filling fraction $n/(2pn+1)$ has a one-to-one correspondence with the IQH states at integer filling factor $n$. 
Only the low relative angular momentum sectors appeared in the Hamiltonian, so we expected that the interaction must be short ranged and that the qualitative results obtained in the disk and spherical geometries should extend to other geometries as well. The interaction presented is not diagonal in position representation and therefore usual approaches to mapping the Hamiltonian from disk geometry to torus or cylinder geometry do not work. Nevertheless, we could construct a Hamiltonian that is motivated by the disk Hamiltonian and has qualitatively the same structure.

The Hamiltonian can be interpreted as that of a multilayer model where different layers have different chemical potentials but each layer is treated as a different LL of same particles. The eigenfunctions of the Hamiltonian when written in the momentum Fock space is then similar to that of a multilayer model. The real space wave functions for multi-Landau level eigenfunctions can be written in a compact form on the disk geometry. 

We showed that the structure of this wave function generalizes in a natural way to the cylinder geometry but not to the torus or spherical geometry. On the torus geometry, we showed that the generalization fails to preserve the right boundary condition. However we could construct the low energy QP excitations of the Laughlin $1/(2p+1)$ state in the spherical geometry (Eq.~\eqref{SphereAltAnsatz}); by generalizing a simplified form (Eq.~\eqref{AlterSolDisk})  of the general eigenfunction on the disk. On the disk, cylinder and spherical geometry we could verify the eigenfunction by comparing with the numerical (ED) results. This wave function when generalized to the torus geometry produces a wave function (Eq.~\eqref{correctWF}) with the correct boundary conditions and expected total kinetic energy. We conjecture that this is also an eigenfunction of the full interacting Hamiltonian. Explicit verification of the result is challenging due to difficulty in explicit evaluation of the wave function.

The model interaction captures some key features of the FQH phases and excitations at the Jain sequence filling fractions and produces exact eigenfunctions with wave functions closely similar in structure to the CF excitations.
We could ask if a similar model interaction can be written which describes more complex FQH liquids. Interestingly the ideas can be generalized, as shown in Ref \onlinecite{Koji22}, to the case of the Moore Read states and allows construction of exact low energy eigenfunctions analogous to the structure of the bipartite composite fermion excitations.\cite{Sreejith11,PhysRevLett.107.086806,Sreejith13} Degeneracy on the torus geometry of the Moore Read states have a non-Abelian component in addition to what is expected from the $q$-fold degeneracy due to the center of mass translations. It is interesting ask how this degeneracy will be manifested in a torus geometry generalization of the results in Ref. \onlinecite{Koji22}. 

\begin{acknowledgments}
G. J. S. acknowledges financial support from DST-SERB (India) Grant No. ECR/2018/001781 and a joint grant from IISER-Pune CNRS. A.A. is supported by SRF-CSIR
(India), Grant No. 09/936(0220)/2019-EMR-I. S.P. acknowledges support by the U.S. Department of Energy, Office of Basic Energy Sciences, under Grant No. DE-SC-0005042, by the Leverhulme Trust Research Leadership Award RL-2019-015 and by EPSRC grant EP/R020612/1.
We thank JK Jain for useful discussions and National Supercomputing Mission
(NSM) for providing computing resources of `PARAM Brahma' at IISER Pune, which is implemented by C-DAC and supported by the Ministry of Electronics and Information Technology (MeitY) and Department of Science and Technology (DST), Government of India.
\end{acknowledgments}

\bibliographystyle{apsrev}
\bibliography{biblio_torus.bib}

\widetext

\begin{center}
	\textbf{\large Appendix}
\end{center}

\setcounter{equation}{0}
\setcounter{table}{0}
\setcounter{page}{1}
\setcounter{section}{0}
\makeatletter
\renewcommand{\thesection}{\Alph{section}}
\renewcommand{\theequation}{A\arabic{equation}}
\renewcommand{\thefigure}{\arabic{figure}}
\renewcommand{\bibnumfmt}[1]{[A#1]}
\renewcommand{\citenumfont}[1]{A#1}

\section{Spectra on torus geometry for other FQH states} \label{AdditionalFigures}
\begin{figure}[h!]
	\includegraphics[width=0.5\columnwidth]{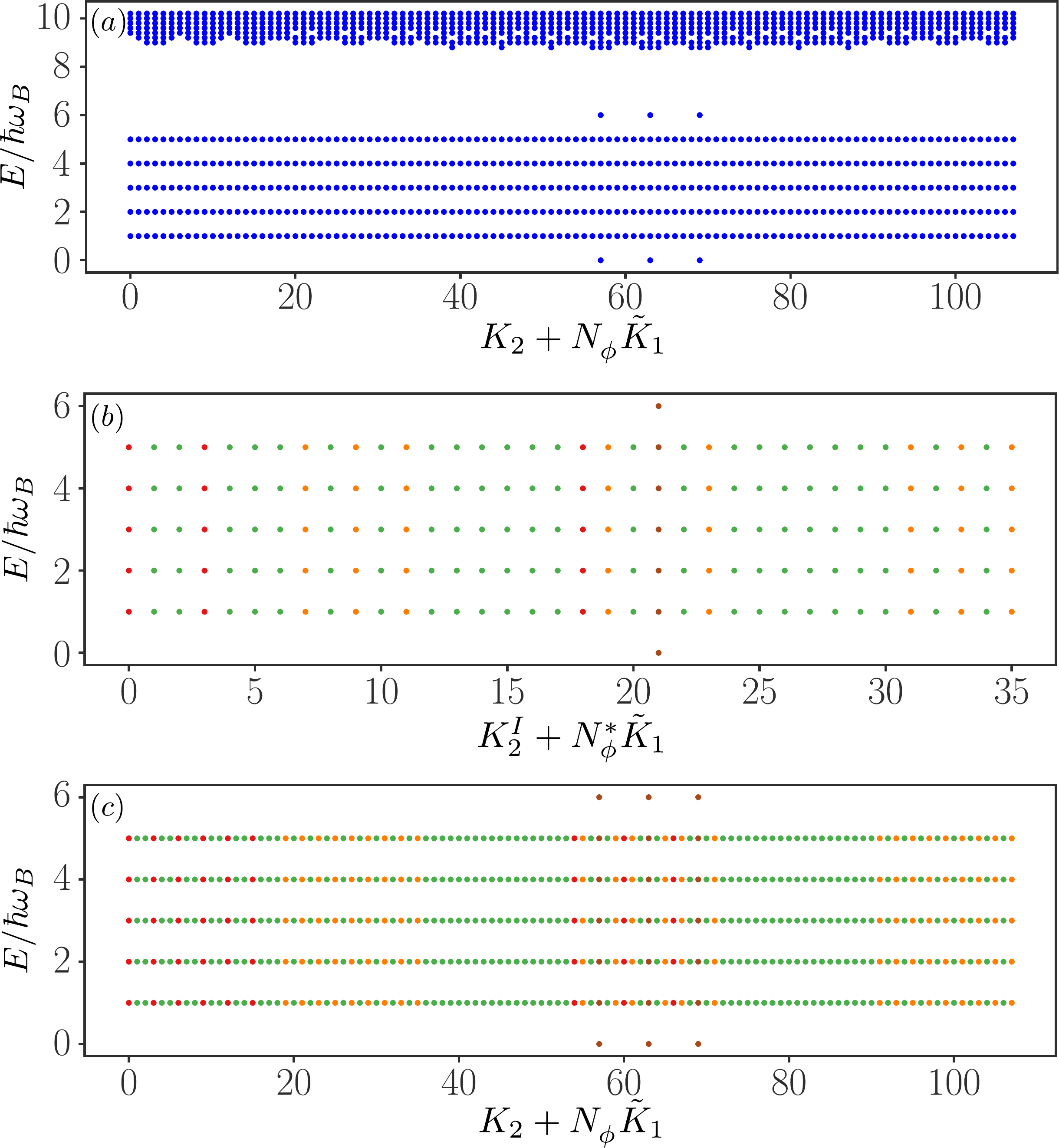}
	\caption{
	In panel $(a)$, we show the spectrum of model interaction for system on the torus with the configuration $(\np,N_e)=(18,6)$ corresponding to filling fraction $\nu=1/3$, where the states are labeled by a pair of quantum numbers $(\tilde{K}_1,K_2)$ along the $x$-axis and the $y$-axis represents their energy which is rescaled such that $\hbar\omega_B\rightarrow 1$. In this panel, we see that the spectrum of the interacting system (FQH) has a clear gap, which separates the spectrum of the  ZIE eigenspace of the interaction, from the finite interaction energy states. The states with finite interaction energy  are higher in the spectra, and only a few of them are visible in the given energy range. These states do not map to the spectra of non-interacting (IQH) system and hence are not of our interest. In panel $(b)$, we show the spectrum of IQH system at integer filling factor $\nu^{*}=1$ for configuration $(\np^{*},N_e)=(6,6)$. The states in a given $(\tilde{K}_1, {K}^{I}_2)$-sector of the IQH spectra are color coded for each unique degeneracy pattern. For instance, the eigenfunctions in red have a degeneracy of $(1,6,12,6,1)$ from low to high energy bands. For system with $N_e/\np=p/q$ where $p,q$ are coprime, each $(\tilde{K}_1, {K}^{I}_2)$-sector of IQH system is mapped to $q$ different sectors of FQH, such that  $(\tilde{K}_1,K_2)=[\tilde{K}_1,K^I_2+r \times \text{gcd}(N_e,\np)]$ where $r=0,1,\dots,q-1$. This threefold multiplicity in FQH spectrum relative to the IQH spectra is demonstrated in the panel $(c)$.
	}
	\label{figA1}
\end{figure}

\begin{figure}[h]
	\includegraphics[width=.25\columnwidth]{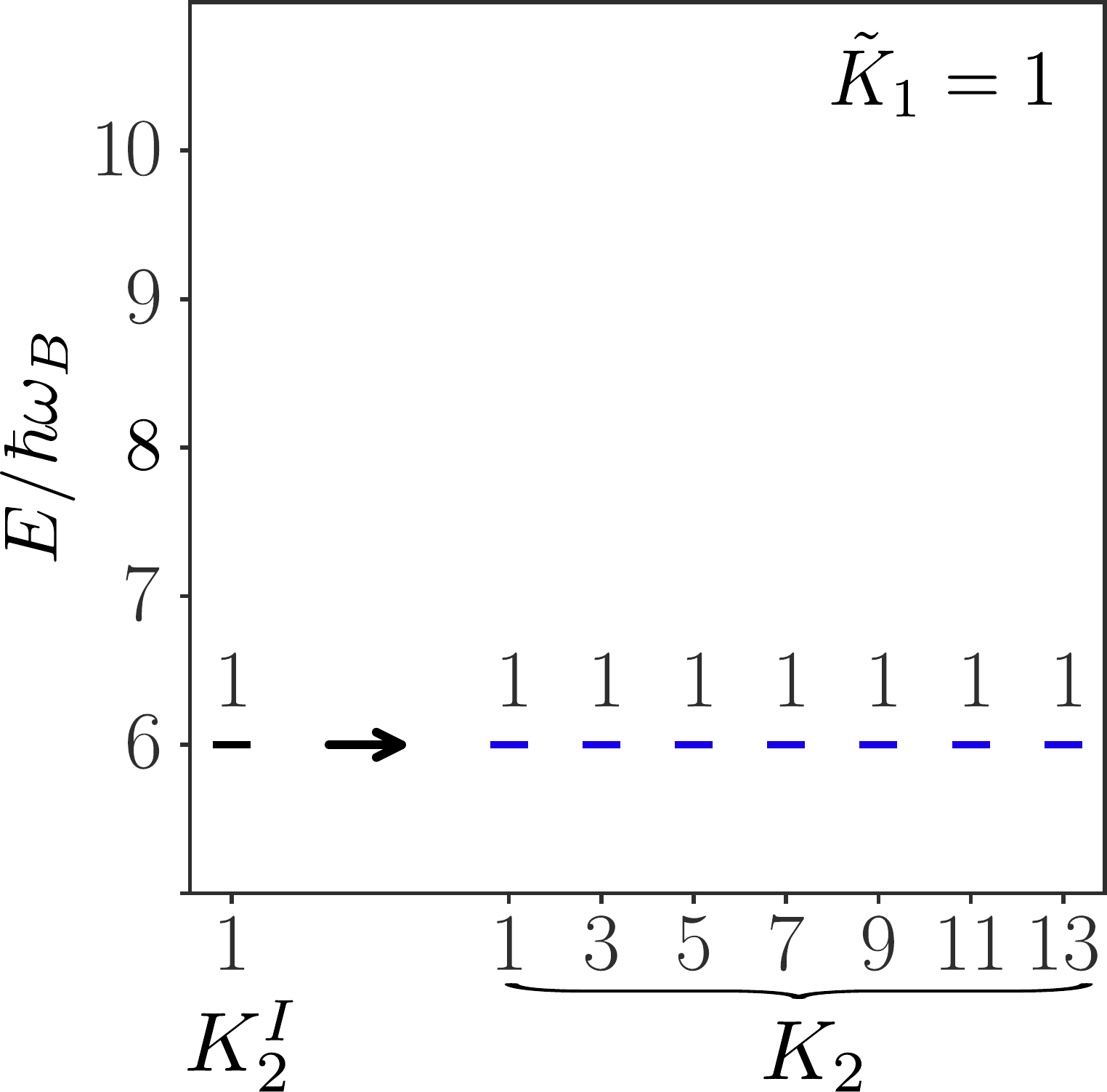}
	\caption{
		Map for low-energy spectra of non-interacting system (black) for $(\np^{*},N_e)=(2,6)$, at $\nu^{*}=3$ to the corresponding spectra of interacting (blue) system for $(\np,N_e)=(14,6)$ at $\nu=3/7$. The map clearly shows a $7$-fold degeneracy. Since our calculations are restricted to lowest $3$ LLs, only $3/7$ ground states are present in the spectra.
	}
	\label{figA2}
\end{figure}

Figure~\ref{figA1} presents the spectrum of model interaction on torus at filling fraction $\nu=\frac{1}{3}$. In Fig.~\ref{figA2}, we show the IQH-FQH mapping for the incompressible ground state of model interaction at $\nu=3/7$ which shows the sevenfold multiplicity.

Panels $(a)$ and  $(b)$, show the IQH-FQH map for the spectra of a single QP and QH at FQH filling $\nu=1/3$, respectively. Similarly, panels $(c)$ and $(d)
$, give the maps for a single QP and QH at FQH filling $\nu=2/5$, respectively. Since $\text{gcd}(\np,N_e)=1$ in all of these cases, there is only one degeneracy pattern in the IQH spectra, hence only one map for any representative $(\tilde{K}_1,K^I_2)$-sector of the IQH spectra is sufficient. As the panel shows, in all four cases, each IQH $K^I_2$-sector  maps to $\np$ $K_2$-sectors of FQH. Since $\text{gcd}(\np^*,N_e)=1$ in all of these cases, all $(\tilde{K}_1,K^I_2)$-sectors in IQH spectra have identical degeneracy pattern.\cite{ManyBodyTranslationsTorus} 

\begin{figure}[h!]
	\includegraphics[width=0.4\columnwidth]{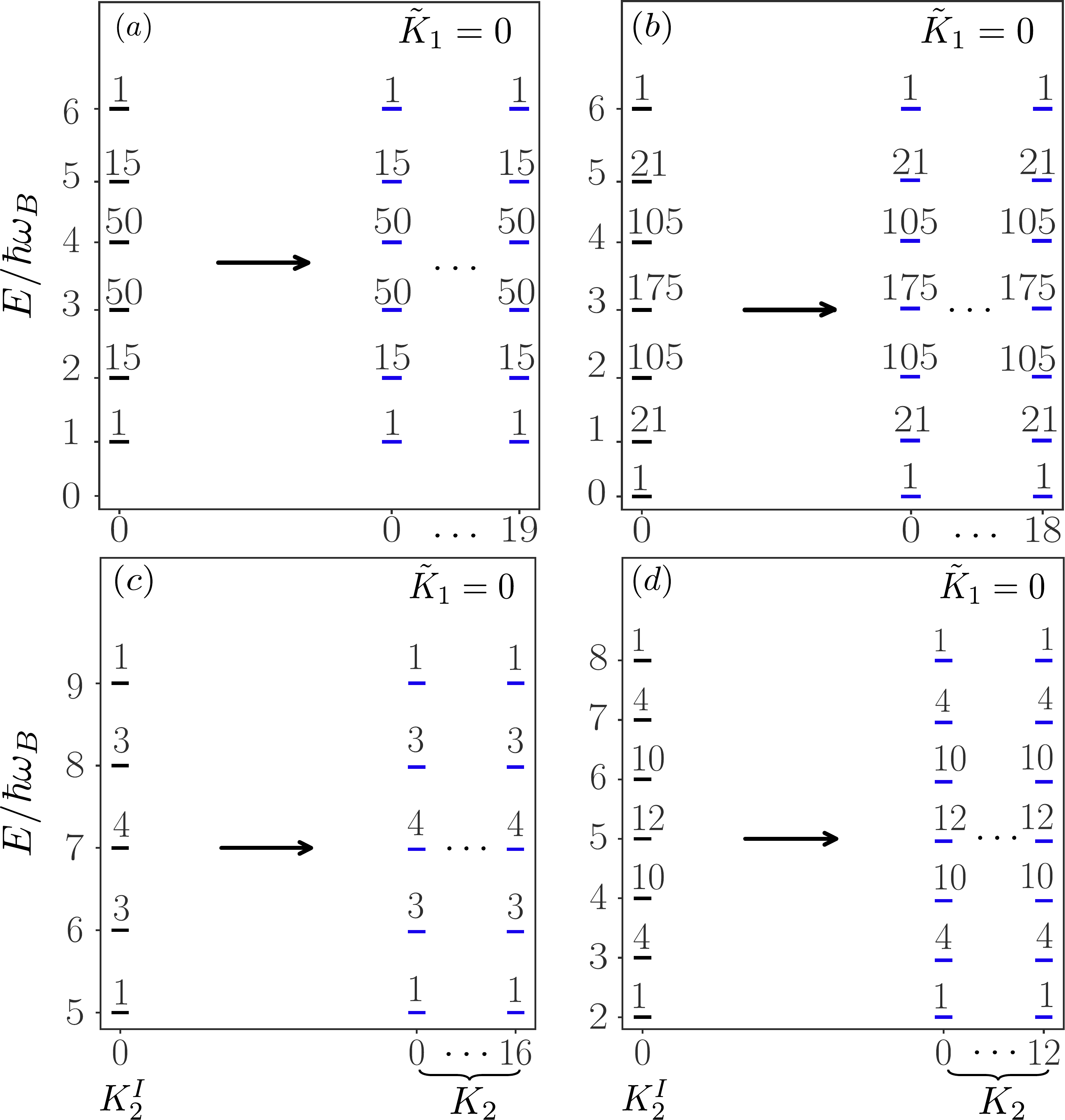}
	\caption{ Plot shows the maps between IQH specta and FQH spectra for single charged excitations (QP/QH) at fillings $\nu=1/3$ and $2/5$. Since $\text{gcd}(\np,N_e)=1$ for all these cases, FQH spectra shows $q=\np$ fold degeneracy. $(a)$ Spectrum for a system hosting a single QP of IQH at $\nu^*=1$ for $N_e=7$ and flux $\np^*=6$ maps to the corresponding FQH spectra at $\nu=1/3$ with flux $\np=20$. The ellipsis $(\dots)$ along the $x$-axis indicate that the FQH system has identical spectra for all intermediate $K_2$ values. Panel $b)$ shows the similar map in system hosting a single QH instead of a QP where IQH spectra for $N_e=6$ and flux $\np^*=7$, at $\nu^*=1$,  maps to the corresponding FQH spectra at $\nu=1/3$ with $\np=19$. Panels $(c)$ and $(d)$ show similar mapping for a single QP and QH between FQH at $\nu=2/5$  with configurations $(\np=17,N_e=7)$ and $(\np=13,N_e=5)$  to the corresponding IQH spectra for configurations $(\np^*=3,N_e=7)$ and $(\np^*=3,N_e=5)$, respectively.
	}
	\label{figA3}
\end{figure}

\section{Calculation of Matrix Elements on Torus geometry} \label{AppToruMatEle}
In this appendix, we show the calculation of matrix elements of the model interaction on the torus. We will restrict to a rectangular torus with $L_1/L_2=\iiota L_x/L_y$ and $L_\Delta=0$.  The matrix elements of an interaction $V(|\bsym{r}|)$ on the torus can be written using its Fourier transform $V(|\bsym{q}|)$  as \begin{align}
V^{n_1 n_2- n_3 n_4}_{j_1 j_2 - j_3 j_4}=&\frac{1}{L_x L_y} \sum_{{q}} V(\bsym{q}) \iint\text{d}\bsym{r}_1 \text{d}\bsym{r}_2 \sbkt{ \phi_{n_1,j_1}^{*}(\bsym{r_1}) \phi_{n_2,j_2}^{*}(\bsym{r_2})-\phi_{n_1,j_1}^{*}(\bsym{r_2}) \phi_{n_2,j_2}^{*}(\bsym{r_1})} \nonumber \\
& \times \sbkt{ \phi_{n_3,j_3}(\bsym{r_1}) \phi_{n_4,j_4}(\bsym{r_2})-  \phi_{n_3,j_3}(\bsym{r_2}) \phi_{n_4,j_4}(\bsym{r_1})} e^{{\iiota} \bsym{q}\cdot (\bsym{r}_1-\bsym{r}_2)} \label{matEleFT}
\end{align}
where the form of single-particle orbitals on torus $\phi_{n,j}$, is given in Eq.~\eqref{SingleParticleTorus} and the summation is over reciprocal vectors $\bsym{q}$ of torus lattice vectors, given by $\bsym{q}=2\pi\sbkt{\frac{s}{L_x},\frac{t}{L_y}}$, such that $s, t\in \mathbb{Z}$. We define
\begin{equation}
\bra{n_1,j_1;n_2,j_2}{e^{{\iiota} \bsym{q}\cdot (\bsym{r}_1-\bsym{r}_2)}}\ket{n_3,j_3;n_4,j_4} = \iint\text{d}\bsym{r}_1 \phi_{n_1,j_1}^{*}(\bsym{r_1}) \phi_{n_2,j_2}^{*}(\bsym{r_2})  e^{{\iiota} \bsym{q}\cdot (\bsym{r}_1-\bsym{r}_2)}  \text{d}\bsym{r}_2 \phi_{n_3,j_3}^{*}(\bsym{r_1}) \phi_{n_4,j_4}^{*}(\bsym{r_2}) \label{definition}
\end{equation}
where $\ket{n_1,j_1;n_2,j_2}$ is a two-particle state for particles in LL $n_1$ and $n_2$ with momentum $j_1$ and $j_2$, respectively. Note that, this two-particle state does not represent an antisymmetrized state.

In Sec.~\ref{SecModelInt}, we provide the details for construction of the model interaction for the torus. The model interaction on the torus is written such that calculation of intra and inter-LL matrix elements uses different forms of $V(|\bsym{q}|)$. For $n$th LL, the intra-LL matrix element is given by
\begin{align}
V^{nn-nn}_{j_1 j_2-j_3 j_4}=& \frac{1}{L_x L_y} \sum_{s,t \in \mathbb{Z}} V(q)\ \sbkt{\mathcal{L}_n\rbkt{\frac{q^2\ell^2}{2}}}^{2}   e^{- \frac{\rbkt{q\ell}^2}{2}}\ \sbkt{  e^{\frac{{\iiota} 2\pi s}{\np} \rbkt{j_1 - j_4}}  -  e^{\frac{{\iiota} 2\pi s}{\np} \rbkt{j_2 - j_4}}  -     e^{\frac{{\iiota} 2\pi s}{\np} \rbkt{j_1 - j_3}}  +   e^{\frac{{\iiota} 2\pi s}{\np} \rbkt{j_2 - j_3}} } \label{A4}
\end{align}
where $\mathcal{L}_n$ is the $n$th Laguerre polynomial and the form of $V(q)=V_1^{nn-nn}(q)$ for intra-LL matrix elements is given in Eq.~\eqref{vq1Intra}.  Using the notation defined in Eq.~\eqref{definition}, inter-LL matrix element for LLs $n$ and $n'$ can be written as
\begin{align}
V^{n n'-n n'}_{j_1 j_2-j_3 j_4} =&  \frac{1}{L_x L_y} \sum_{s,t\in \mathbb{Z}}\ V({q}) \left\{ \bra{n,j_1;n',j_2}{e^{{\iiota} \bsym{q}\cdot (\bsym{r}_1-\bsym{r}_2)}}\ket{n,j_3;n',j_4} + \bra{n',j_2; n,j_1}{e^{{\iiota} \bsym{q}\cdot (\bsym{r}_1-\bsym{r}_2)}}\ket{n',j_4; n,j_3}  
\right.  \nonumber\\        
& \left.   - \bra{n',j_2;n,j_1}{e^{{\iiota} \bsym{q}\cdot (\bsym{r}_1-\bsym{r}_2)}}\ket{n,j_3;n',j_4} - \bra{n,j_1;n',j_2 }{e^{{\iiota} \bsym{q}\cdot (\bsym{r}_1-\bsym{r}_2)}}\ket{n',j_4; n,j_3}  
     \right\}  \label{interLL}
\end{align}
where the explicit form of terms inside the parenthesis  is following
\begin{align}
	\bra{n,j_1;n',j_2}{e^{{\iiota} \bsym{q}\cdot (\bsym{r}_1-\bsym{r}_2)}}\ket{n,j_3;n',j_4}=&  {\mathcal{L}_{n_1}\rbkt{\frac{q^2\ell^2}{2}} \mathcal{L}_{n_2}\rbkt{\frac{q^2\ell^2}{2}}} e^{\frac{{\iiota} 2\pi s}{\np} \rbkt{j_1 - j_4}} \nonumber \\
	\bra{n',j_2;n,j_1}{e^{{\iiota} \bsym{q}\cdot (\bsym{r}_1-\bsym{r}_2)}}\ket{n',j_4;n,j_3}=&  {\mathcal{L}_{n_1}\rbkt{\frac{q^2\ell^2}{2}} \mathcal{L}_{n_2}\rbkt{\frac{q^2\ell^2}{2}}} e^{\frac{{\iiota} 2\pi s}{\np} \rbkt{j_2 - j_3}} \nonumber \\
	\bra{n',j_2;n,j_1}{e^{{\iiota} \bsym{q}\cdot (\bsym{r}_1-\bsym{r}_2)}}\ket{n,j_3;n',j_4}=&  {\frac{\rbkt{q\ell}^{2}}{2}} e^{\frac{{\iiota} 2\pi s}{\np} \rbkt{j_1 - j_3}} \nonumber \\
	\bra{n,j_1;n',j_2}{e^{{\iiota} \bsym{q}\cdot (\bsym{r}_1-\bsym{r}_2)}}\ket{n',j_4; n,j_3}=& {\frac{\rbkt{q\ell}^{2}}{2}} e^{\frac{{\iiota} 2\pi s}{\np} \rbkt{j_2 - j_4}}
\end{align}
For the case of inter-LL term, we use $V(\bsym{q})=V^{n_1n_2-n_3n_4}_0(q)+V^{n_1n_2-n_3n_4}_1(q)$. 

Putting the corresponding form of V(q) given in Sec.~\ref{SecModelInt}, intra and inter-LL matrix elements are given by
\begin{align}
V^{n n-n n}_{j_1 j_2-j_3 j_4} =&  \frac{1}{L_x L_y} \sum_{s,t\in \mathbb{Z}} e^{ -\frac{{q}^2}{2}}\ \mathcal{L}_1(q^2\ell^2) \rbkt{   {e^{\frac{{\iiota} 2\pi s}{\np} \rbkt{j_1 - j_4}} + e^{\frac{{\iiota} 2\pi s}{\np} \rbkt{j_2 - j_3}}}  -  { e^{\frac{{\iiota} 2\pi s}{\np} \rbkt{j_1 - j_3}} -  e^{\frac{{\iiota} 2\pi s}{\np} \rbkt{j_2 - j_4}}} }   
\end{align}
and 
\begin{align}
V^{n n'-n n'}_{j_1 j_2-j_3 j_4} =&  \frac{1}{L_x L_y} \sum_{s,t\in \mathbb{Z}} e^{ -\frac{{q}^2}{2}}\  \sbkt{   (\mathcal{L}_1(q^2\ell^2) + \mathcal{L}_0(q^2\ell^2)) \rbkt{e^{\frac{{\iiota} 2\pi s}{\np} \rbkt{j_1 - j_4}} + e^{\frac{{\iiota} 2\pi s}{\np} \rbkt{j_2 - j_3}}} \right. \nonumber\\
	& \left. -   (\mathcal{L}_1(q^2\ell^2) - \mathcal{L}_0(q^2\ell^2)) \rbkt{ e^{\frac{{\iiota} 2\pi s}{\np} \rbkt{j_1 - j_3}} +  e^{\frac{{\iiota} 2\pi s}{\np} \rbkt{j_2 - j_4}}} }   
\end{align}
Both intra and inter-LL matrix elements are independent of LL-indices.

\section{Calculation of Matrix Elements on Cylinder geometry}

\begin{figure}[h!]
	\centering
	\includegraphics[width=2in]{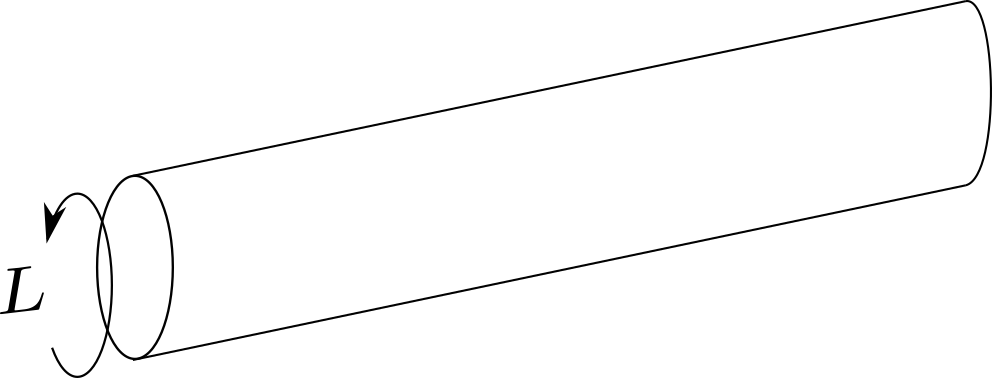}
	\caption{Cylinder geometry}
	\label{cylinder}
\end{figure}

For a cylinder given in Fig. \ref{cylinder}, single-particle orbitals are given by
\begin{equation}\label{cylinderSPWF}
\phi_{n,k}(\bsym{r}) = \frac{1}{\sqrt{2^n n! \sqrt{\pi}L}}\text{exp}\sbkt{-{\iiota}\frac{2\pi k}{L}y}\text{exp}\sbkt{-\frac{1}{2}\rbkt{\frac{x}{\ell}-\frac{2\pi \ell}{L}k}^2}H_n\sbkt{\frac{2\pi \ell}{L}k -\frac{x}{\ell}}
\end{equation}
where $L$ is the length of circumference. Length of the cylinder fixed by putting a cut-off on $k$ values such that it can only take $\np$ consecutive values in each LL with $k_{\rm min}=0$ and $k_{\rm min}=\np-1$.
The matrix-elements for cylinder are calculated using
\begin{equation}\label{cyl_vhat}
{V} = \frac{1}{(2\pi L)}\sum_{m}\int \text{d}q\ V(\bsym{q}) e^{{\iiota} q (\hat{x}_1 - \hat{x}_2)} e^{{\iiota} \frac{2\pi m}{L} (\hat{y}_1 - \hat{y}_2)}
\end{equation}
For calculation of cylinder matrix elements, we use the same form of $V(q)$ which was used in the case of torus geometry.

\section{Ansatz for Quasi-particles of $\nu=1/3$ on Disk} \label{Ansatz1by3QPDisc}
Even though the exact eigenfunctions for the model interaction on the disk geometry can be written in a compact form, given in Eq.~\eqref{diskansatz}, these do not immediately generalize to the case of the spherical or torus geometry. In this appendix we describe a form of the ansatz that is equivalent to Eq.~\eqref{diskansatz} for the special case of quasiparticles of 1/3. The form presented here has the advantage of generalizing to other geometries. In this appendix, we restrict to the case where all particles are in the lowest two LLs \ie $n=0,\,1$ as is appropriate when considering quasiparticles of 1/3.

The state with angular momentum $m$ in $n$th LL has the form
\begin{equation}
	\eta_{n,m}(z,\bar{z})=F_{n,m}(z,\bar{z})\,e^{-|z|^2/4\ell^2}
\end{equation}
where $F_{n,m}(z,\bar{z})$ is a polynomial of $z$ and $\bar{z}$. Highest power of $\bar{z}$ in $F_{n,m}(z,\bar{z})$ is equal to the LL-index $n$. 
The action of the guiding center coordinate $\hat{Z}=z/2-2\ell^2\partial_{\bar{z}}$ on the single particle states $\eta_{n,m}$'s can be reduced to the action of an operator on $F_{n,m}$:
\begin{equation}
\hat{Z}\eta_{n,m}(z,\bar{z})= \rbkt{z/2-2\ell^2\partial_{\bar{z}}} F_{n,m}(z,\bar{z})\,e^{-|z|^2/4\ell^2} =e^{-|z|^2/4\ell^2} \rbkt{z-2\ell^2\partial_{\bar{z}}} F_{n,m}(z,\bar{z})
\end{equation}
In the remaining calculations we will omit the exponential factor from the expressions.

We will now consider the state describing $N$ QPs of $1/3$ given by, $\Psi^{\rm N-QPs}_{\nu=1/3} =  \mathcal{J}^{2}(\{ \hat{Z}_i \})\,\Phi^{\rm N-QPs}_{1} $, where the $\Phi^{\rm N-QPs}_{1}$ contains $N$ particles in LL1.  Any ansatz state is called a proper state when for each occupied Landau orbital in the Slater determinant $\Phi_{\nu^*}$, with LL index $n$ and momentum state $k$, $k$th momentum Landau orbitals in all lower LLs are also filled.\cite{Pu2017} Hence, quasiparticle states are proper states. In a determinant $\Phi^{\rm N-QPs}_{1}$ corresponding to a proper state, the orbitals in the second Landau can be written as $F_{1,m}(z_i,\bar{z}_i)=z_i^m\bar{z}_i$ for all $\textit{i}$, without affecting the Slater determinant $\Phi^{\rm N-QPs}_{1}$. Hereafter we will use this as the definition of $F_{1,m}(z_i,\bar{z}_i)$. The remaining terms in $F_{1,m}$ do not contribute to the determinant.

Since all particles in $\Phi$ are in the lowest two LLs $\Phi^{\rm N-QPs}_{1}$ will at most be linear in $\bar{z}_i$'s, for each $i$. 
This implies that we only need to expand the Jastrow factor $\mathcal{J}^{2}(\{ \hat{Z}\})$ up to linear terms in $\partial_{\bar{z}}$'s:
\begin{multline}
\mathcal{J}^{2}(\{ \hat{Z} \})\Phi^{\rm N-QPs}_{1}=\prod_{i<j} (\hat{Z}_i - \hat{Z}_j)^{2}\Phi^{\rm N-QPs}_{1}
=\prod_{i}\sbkt{\mathcal{J}^2(\{z\}) - \partial_{z_i} \mathcal{J}^2(\{z\})\,\rbkt{2\ell^2\partial_{\bar{z}_is}}}\Phi^{\rm N-QPs}_{1}=\\
=\prod_{i}\sbkt{ 1 - \partial_{z_i} \,(2\ell^2\partial_{\bar{z}_i})}\Phi^{\rm N-QPs}_{1}\mathcal{J}^2(\{z\})
\end{multline}
Here $\mathcal{J}(\{z\})=\prod_{i<j} (z_i - z_j)$ is the Jastrow factor of normal position coordinates. Note that in the last expression the derivatives $\partial_{z_i} $'s  act only on the Jastrow factor $\mathcal{J}^2(\{z\})$ and $\partial_{\bar{z}_i}$ acts only on the Slater determinant $\Phi^{\rm N-QPs}_{1}$.

The above expression acts trivially on lowest Landau level states of the Slater Determinant:
\begin{equation}
\sbkt{ 1 - 2\ell^2  \partial_{z_i} \,\partial_{\bar{z}_i}}  z_i^{m_i}\mathcal{J}^2(\{z\})  = z_i^{m_i}\mathcal{J}^2(\{z\})
\end{equation}
When there are states from the second LL in the Slater determinant it acts as
\begin{equation}
\sbkt{ 1 - 2\ell^2  \partial_{z_i} \,\partial_{\bar{z}}}  \bar{z}_i z_i^{m_i}\mathcal{J}^2(\{z\})  =  \sbkt{ \bar{z}_i - 2\ell^2  \partial_{{z}_i}} z_i^{m_i}  \mathcal{J}^2(\{z\}) = P_{\rm LL1} z_i^m \bar{z}_i\mathcal{J}^2(\{z\})
\end{equation}
where $\mathcal{P}_{\rm LL1}=\mathbb{I}-\mathcal{P}_{\rm LLL}$ is the projection to the second LL.

Combining these results, the ansatz simplifies to the following expression for the case of $N$ quasiparticles of $1/3$.
\begin{equation}
 \mathcal{J}^{2}(\{ \hat{Z}_i \})\,\Phi^{\rm N-QPs}_{1} = \hat{\Phi}^{\rm N-QPs}_{1} \mathcal{J}^{2}(\{ z_i \})
\end{equation}
where the Slater determinant $\hat{\Phi}^{\rm N-QPs}_{1}$ is constructed by replacing all LL1 Landau orbitals $F_{1,m}(z,\bar{z})$, inside $\Phi^{\rm N-QPs}_{1}$ by $F_{1,m}(z,\bar{z})- \hat{F}_{1,m}(z,\bar{z}\rightarrow 2\ell^2 \partial_z)$.  Here the operator $\hat{F}_{1,m}(z,\bar{z}\rightarrow 2\ell^2 \partial_z)$ represents LLL projection of the LL1 Landau orbital $F_{1,m}(z,\bar{z})$ constructed by replacing $\bar{z}\rightarrow 2\ell^2 \partial_z$. A normal ordering is required  such that all $\bar{z}$ are moved to the left before making the replacement $\bar{z}\rightarrow 2\ell^2 \partial_z$ where it is understood that the derivatives do not act on the exponentials.\cite{Jain07} The exact eigenfunction given in Eq.~\eqref{diskansatz} can be rewritten in this way for neutral excitations of $1/3$ as well, as long as the Slater determinant state $\Phi^{\alpha}_{1}$ is a proper state.

\section{Details of periodicity checks for torus ansatz} \label{periodicityChecks}
In this appendix, we discuss the details of periodic boundary condition checks of both torus ansatz, given in Sec.~\ref{IncorrectAnsatz} and \ref{correctAnsatz}. We will use following properties of Jacobi theta functions, given by
\begin{gather}
\rtheta{a}{b}{z\pm1}{\tau} = e^{\pm\iiota 2\pi a}\rtheta{a}{b}{z}{\tau}\nonumber
\\[5pt]
\rtheta{a}{b}{z\pm\tau}{\tau} = e^{-\iiota \pi [\tau \pm 2(z+b)]}\rtheta{a}{b}{z}{\tau} 	\label{app:theta}
\end{gather}
where for the torus given torus in Fig.~\ref{fig::2.1}, $\tau=-L_1/L_2$.

First, we will discuss periodicity checks for ansatz given in Sec.~\ref{IncorrectAnsatz}. The action of  magnetic translation operators (MTOs) $t_i(L_1)$ and $t_i(L_2)$ the exponential factor ${e^{-\frac{\sum_i z^2_i+|z_i|^2 }{2\ell^2}}}$ is given by
\begin{gather}
t_i(L_2){e^{-\frac{\sum_i z^2_i+|z_i|^2 }{2\ell^2}}}= {e^{-\frac{\sum_i z^2_i+|z_i|^2 }{2\ell^2}}} \\
t_i(L_1){e^{-\frac{\sum_i z^2_i+|z_i|^2 }{2\ell^2}}}= e^{\iiota \pi \np \rbkt{\tau-\frac{2z_i}{L_2}} }{e^{-\frac{\sum_i z^2_i+|z_i|^2 }{2\ell^2}}}
\end{gather}
where we use Eq.~\ref{symmMTO} to write the MTOs as a phase times corresponding normal translation operators. No that the phase term is taken into account, we only need to calculate the action of normal translation operators $T_i(L_1)$ and $T_i(L_2)$, on the remaining parts of the ansatz. Action of these on single-particle wave fucntions (Eq~\eqref{singleParticleOrbitals}) on the torus is given by
\begin{gather}
T_i(L_2) f_0^{k}(z_i) = \rtheta{\frac{k}{\np^*} +\frac{\theta_2}{2\pi \np^*}}{\frac{\theta_1}{2\pi}}{\frac{\np^* z_i}{L_2} +\np^*}{\np^*\tau} = e^{\iiota \theta_2} f_0^{k}(z_i) \label{ATL2_1}\\
T_i(L_1) f_0^{k}(z_i) = \rtheta{\frac{k}{\np^*} +\frac{\theta_2}{2\pi \np^*}}{\frac{\theta_1}{2\pi}}{\frac{\np^* z_i}{L_2} -\np^*\tau}{\np^*\tau} = e^{\iiota \theta_1} e^{-\iiota\pi\np^*\rbkt{\tau-\frac{2z_i}{L_2}}} f_0^{k}(z_i)  \label{ATL1_1}
\end{gather}
and similarly, we have
\begin{gather}
T_i(L_2) f_1^{k}(z_i,\bar{z}_i) = \sqrt{2}{\ell^{*}}\sbkt{\frac{\bar{z}_i+z_i}{2(\ell^*)^{2}} -\partial_{z_i}} e^{\iiota \theta_2} f^{k}_0(z_i) = e^{\iiota \theta_2} f_1^{k}(z_i,\bar{z}_i) \label{ATL1_2}\\
T_i(L_1) f_1^{k}(z_i,\bar{z}_i) = \sqrt{2}{\ell^{*}}\sbkt{\frac{\bar{z}_i+z_i}{2(\ell^*)^{2}} +\frac{L_x}{(\ell^*)^2} -\partial_{z_i}} e^{\iiota \theta_1} e^{-\iiota\pi\np^*\rbkt{\tau-\frac{2z_i}{L2}}} f_0^{k}(z_i)  = e^{\iiota \theta_1} e^{-\iiota\pi\np^*\rbkt{\tau-\frac{2z_i}{L_2}}} f_1^{k}(z_i,\bar{z}_i) \label{ATL2_2}
\end{gather}
Since the Slater determinant $\Phi_{\nu^*}$ consists of LLL states $f_0^k(z)$ and LL1 $f_1^k(z,\bar{z})$ only, the action of $T_i(L_1)$ and $T_i(L_2)$ on $\Phi_{\nu^*}$ 
is identical. Putting together Eq.~\eqref{ATL2_1} and \eqref{ATL2_2}, gives up Eq.~\eqref{WrongWF1} and Eq.~\eqref{ATL1_1} and \eqref{ATL1_2} combine to give  Eq.~\eqref{lastEqFail}.

For the ansatz given in Sec.~\ref{correctAnsatz}, the action of  translation  $T_i(L_2)$ is trivial. The action of $T_i(L_1)$ on the Jastrow factor is given by
\begin{align}
	T_i(L_1)\mathcal{J}^2 =&\,  e^{-2\iiota\pi(N_e-1)\tau}e^{-4\iiota\pi (Z_{\rm cm}-N_e z_i)/L_2}\mathcal{J}^2
\end{align}
When put together with Eq.~\eqref{ATL1_1}, it gives Eq.~\eqref{tl23}. Similarly the action of $T_i(L_1)$ on $ \hat{g}_1^{q_j}(z_i)\mathcal{J}^2$ is given by
\begin{align}
T_i(L_1)\hat{g}_1^{q}(z_i)\mathcal{J}^2=& \frac{\sqrt{2}\np^*\ell^*}{\np} \sbkt{ \frac{\bar{z}_i+z_i}{2\ell^{2}}f^{q}_0(z_i+L_1)+ \frac{L_x}{\ell^{2}}f^{q}_0(z_i) - \partial_{z_i} f^{q}_0(z_i+L_1) - f^{q}_0(z_i+L_1) \partial_{z_i}} \times \nonumber \\
& \times e^{-2\iiota\pi(N_e-1)\tau}e^{-4\iiota\pi (Z_{\rm cm}-N_e z_i)/L_2}\mathcal{J}^2
\end{align}
By putting the value of Eq.~\ref{ATL1_1} in place of $f^{q}_0(z_i+L_1)$, we get the result in Eq.~\ref{ghatProblem}.

\section{Adiabatic continuity between the neutral excitations of model interaction and LLL Coulomb interaction} \label{NeutralModesAdiabatic}

\begin{figure}[h!]
	\includegraphics[width=0.3\columnwidth]{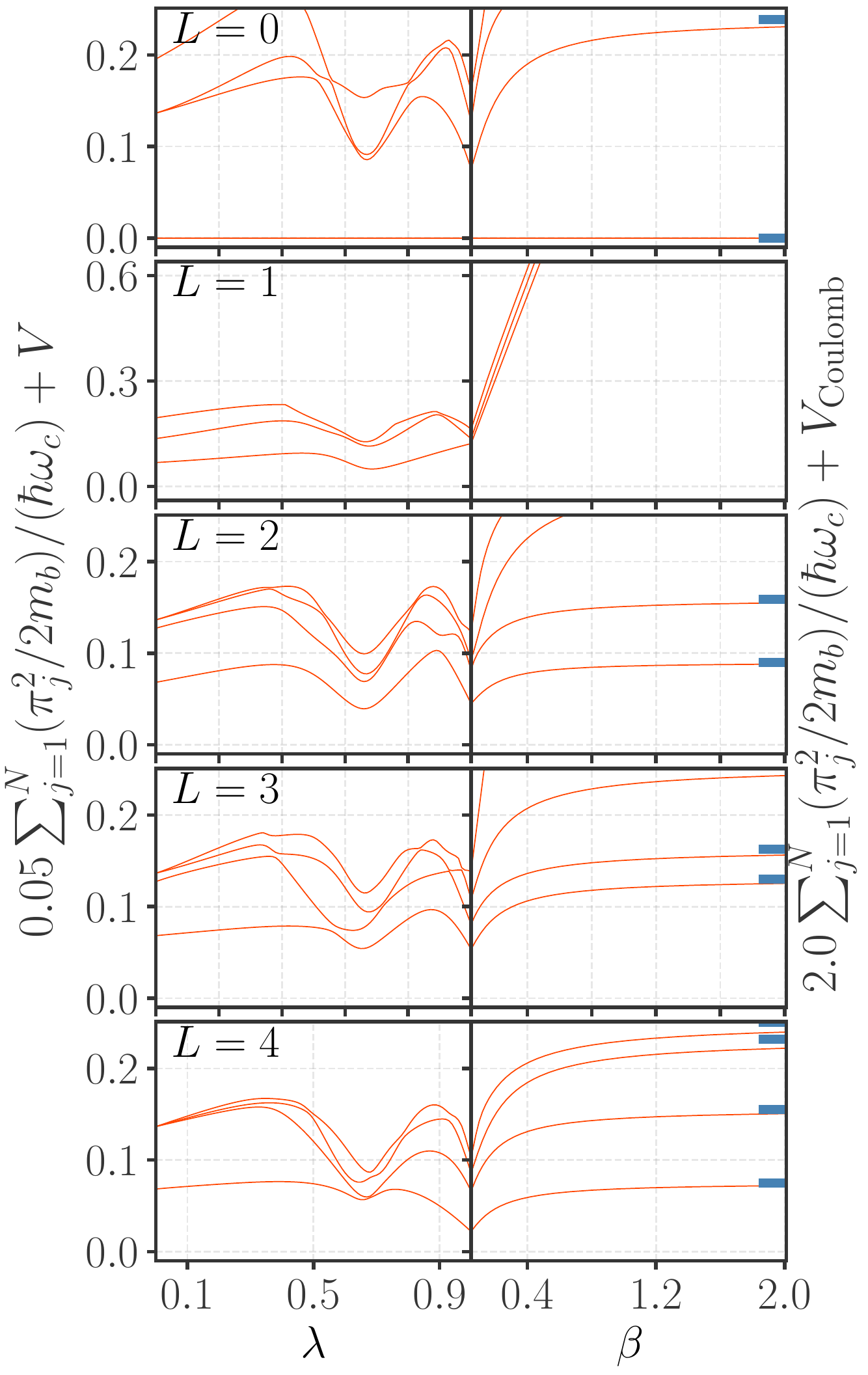}
	\caption{ Demonstration of adiabatic continuity between the ground state and the low energy neutral excitations of the model Hamiltonian and the lowest Landau level Coulomb Hamiltonian for $N=6$ particles at $\nu=2/5$. 
	}
	\label{figA5}
\end{figure}

Figure ~\ref{figA5} shows the study of adiabatic continuity between model and LLL projected Coulomb interaction at filling fraction $\nu=2/5$, where the Hamiltonian is defined as

\begin{equation}
\hat{H}=\beta \sum_{j=1}^N (\hat{\pi}_j^2/2m_b)/(\hbar \omega_c) + (1-\lambda)\hat{V} + \lambda \hat{V}_{\rm Coulomb}
\end{equation}
where $\hat{V}$ is the model interaction of Eq.~\ref{diskModelInter} for sphere, with all its pseudopotentials set to unity, $\hat{V}_{\rm Coulomb}$ is the Coulomb interaction, and all energies are quoted in the units of $e^2/\epsilon \ell$. At each $\lambda$ and $\beta$, the spectrum has been vertically shifted to set the $L=0$ ground state to $0$. The left panels show the evolution of the low-lying eigenstates as $\lambda$ is varied from $0$ to $1$ with $\beta=0.05$. The right panels show the evolution as $\beta$ is changed from $0.05$ to $2.0$ with $\lambda=1.0$. Lowest three LLs are included in this calculation. Different rows indicate the spectra in different angular momentum $(L)$ sectors. Spectra at the leftmost part corresponds to the model interaction, whereas the rightmost spectra are for Coulomb interaction with large cyclotron gap in LLs. The solid blue dashes at the rightmost end show the Coulomb energies in the LLL (\ie for $\lambda=1$ and $\beta=\infty$). Adiabatic continuity for the low-energy states is seen in all cases; the qualitatively different behavior for $L=1$ (note the different energy scale for this row), for which the states are pushed to very high energies, captures the absence of a low-energy neutral mode in the LLL Coulomb spectrum.

\begin{figure}[h!]
	\includegraphics[width=0.3\columnwidth]{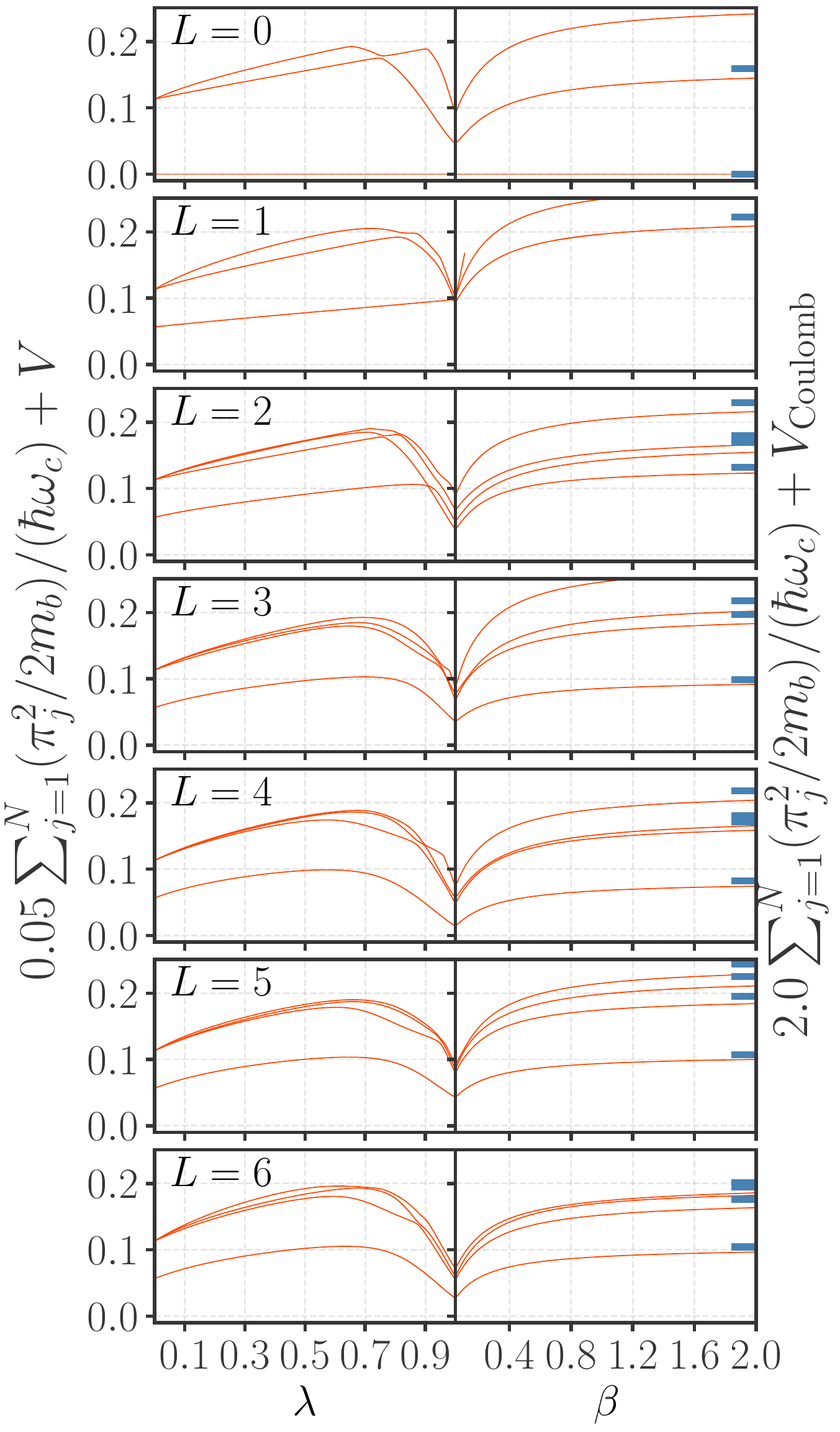}
	\caption{ Demonstration of adiabatic connectivity of low energy eigenstates of our model Hamiltonian and the those of the LLL Coulomb Hamiltonian for a six particle system at $\nu=1/3$. 
	}
	\label{figA6}
\end{figure}

Figure ~\ref{figA6} shows the results of a similar study for filling fraction $\nu=1/3$. The left panel shows the change in the spectra as $\lambda$ changes from $0$ to $1$ keeping $\beta=0.05$ and the right panels show the variation as $\beta$ goes from $0.05$ to $2$. Different rows shows the spectra in different $L$ sectors. All energies are relative to the ground state of $L=0$ sector.
Here as well, in the $L=1$ sector, one of the three states from the left-hand side is projected out to high energy as the cyclotron gap $\beta$ is increased.
The ground state of the model Hamiltonian adiabatically connects to ground state for LLL Coulomb Hamiltonian. The same is true from neutral excitations, with the exception of level-crossing at $L=2$.

\end{document}